\documentclass[journal]{IEEEtran}
\usepackage{cite}

\ifCLASSINFOpdf
  \usepackage[pdftex]{graphicx}
  \graphicspath{{./img/}}
  \DeclareGraphicsExtensions{.pdf,.jpeg,.png}
\else
  \DeclareGraphicsExtensions{.eps}
\fi
\usepackage{amsmath,amssymb}
\usepackage{siunitx}
\usepackage{bm}
\usepackage{mathtools,cuted}
\usepackage{empheq}
\usepackage{tikz,pgfplots}
\usetikzlibrary{shapes, arrows, chains, calc, positioning}
\pgfplotsset{compat=newest,
  	grid style={solid, lightgray},
      	unbounded coords=jump, 
      	grid=major, 
      	width=0.4\textwidth,
      	height=0.3\textwidth,
	legend style={draw=none},
	enlargelimits=false}

\newcommand{\ML}{M_{L}}
\newcommand{\MR}{M_{R}}
\newcommand{\links}{_{L}}
\newcommand{\rechts}{_{R}}
\newcommand{\Sx}{s_{x}}

\newcommand{\Su}{s_{u}}

\newcommand{\YL}{y\links}
\newcommand{\YR}{y\rechts}

\newcommand{\ZL}{z\links}
\newcommand{\ZR}{z\rechts}

\newcommand{\wL}{\mathbf{w}_{L}}
\newcommand{\wR}{\mathbf{w}_{R}}
\newcommand{\wx}{\mathbf{w}_{x}}
\newcommand{\wxl}{\mathbf{w}_{x,L}}
\newcommand{\wxr}{\mathbf{w}_{x,R}}
\newcommand{\wu}{\mathbf{w}_{u}}
\newcommand{\wul}{\mathbf{w}_{u,L}}
\newcommand{\wur}{\mathbf{w}_{u,R}}

\newcommand{\wLblcmv}{\mathbf{w}_{\mathrm{BLCMV},L}}

\newcommand{\wLblcmvn}{\mathbf{w}_{\mathrm{BLCMV-N},L}}
\newcommand{\wRblcmvn}{\mathbf{w}_{\mathrm{BLCMV-N},R}}

\newcommand{\vecy}{\mathbf{y}}
\newcommand{\vecx}{\mathbf{x}}
\newcommand{\vecn}{\mathbf{n}}
\newcommand{\vecu}{\mathbf{u}}
\newcommand{\vecv}{\mathbf{v}}
\newcommand{\veca}{\mathbf{a}}

\newcommand{\vecb}{\mathbf{b}}

\newcommand{\veceL}{\mathbf{e}\links}
\newcommand{\veceR}{\mathbf{e}\rechts}
\newcommand{\vecgl}{\mathbf{g}\links}

\newcommand{\vecgxl}{\mathbf{g}_{x,L}}

\newcommand{\vecgul}{\mathbf{g}_{u,L}}

\newcommand{\matRx}{\mathbf{R}_{x}}
\newcommand{\matRu}{\mathbf{R}_{u}}
\newcommand{\matRn}{\mathbf{R}_{n}}
\newcommand{\matRv}{\mathbf{R}_{v}}
\newcommand{\matRy}{\mathbf{R}_{y}}

\newcommand{\matC}{\mathbf{C}}

\newcommand{\ga}{\gamma_{a}}
\newcommand{\gb}{\gamma_{b}}
\newcommand{\gab}{\gamma_{ab}}

\newcommand{\SNRinL}{\mathrm{SNR}\links^{\mathrm{in}}}
\newcommand{\SNRinR}{\mathrm{SNR}\rechts^{\mathrm{in}}}
\newcommand{\SNRoutL}{\mathrm{SNR}\links^{\mathrm{out}}}
\newcommand{\SNRoutR}{\mathrm{SNR}\rechts^{\mathrm{out}}}
\newcommand{\SNRoutLmvdr}{\mathrm{SNR}_{\mathrm{BMVDR},L}^{\mathrm{out}}}

\newcommand{\SNRoutLmvdrn}{\mathrm{SNR}_{\mathrm{BMVDR-N},L}^{\mathrm{out}}}

\newcommand{\SIRoutLlcmv}{\mathrm{SIR}_{\mathrm{BLCMV},L}^{\mathrm{out}}}

\newcommand{\SNRoutLlcmvn}{\mathrm{SNR}_{\mathrm{BLCMV-N},L}^{\mathrm{out}}}

\newcommand{\SIRoutLlcmvn}{\mathrm{SIR}_{\mathrm{BLCMV-N},L}^{\mathrm{out}}}

\newcommand{\SIRinL}{\mathrm{SIR}\links^{\mathrm{in}}}
\newcommand{\SIRinR}{\mathrm{SIR}\rechts^{\mathrm{in}}}
\newcommand{\SIRoutL}{\mathrm{SIR}\links^{\mathrm{out}}}
\newcommand{\SIRoutR}{\mathrm{SIR}\rechts^{\mathrm{out}}}

\begin{document}
%
\title{{Binaural LCMV Beamforming with\\ Partial Noise Estimation}}
%
%
%

\author{Nico~G\"o{\ss}ling,~\IEEEmembership{Student~Member,~IEEE,}
        Elior~Hadad,~\IEEEmembership{Member,~IEEE,}
        Sharon~Gannot,~\IEEEmembership{Senior~Member,~IEEE}
        and~Simon~Doclo,~\IEEEmembership{Senior~Member,~IEEE}%
		\thanks{This work was funded by the Deutsche Forschungsgemeinschaft (DFG, German Research Foundation) - Project ID 352015383 (SFB 1330 B2) and Project ID 390895286 (EXC 2177/1) and the Israeli Ministry of Science and Technology, \#88962, 2019.}%
		\thanks{N. G\"o{\ss}ling and S. Doclo are with the Department of Medical Physics and Acoustics and the Cluster of Excellence Hearing4all, University of Oldenburg, 26111 Oldenburg, Germany (e-mail: nico.goessling@uol.de; simon.doclo@uol.de).}%
		\thanks{E. Hadad and S. Gannot are with the Faculty of Engineering, Bar-Ilan University, Ramat-Gan, 5290002, Israel (e-mail: elior.hadad@biu.ac.il; sharon.gannot@biu.ac.il).}%
}

\maketitle

\begin{abstract}
Besides reducing undesired sources, i.e., interfering sources and background noise, another important objective of a binaural beamforming algorithm is to preserve the spatial impression of the acoustic scene, which can be achieved by preserving the binaural cues of all sound sources.
While the binaural minimum variance distortionless response (BMVDR) beamformer provides a good noise reduction performance and preserves the binaural cues of the desired source, it does not allow to control the reduction of the interfering sources and distorts the binaural cues of the interfering sources and the background noise.
Hence, several extensions have been proposed.
First, the binaural linearly constrained minimum variance (BLCMV) beamformer uses additional constraints, enabling to control the reduction of the interfering sources while preserving their binaural cues.
Second, the BMVDR with partial noise estimation (BMVDR-N) mixes the output signals of the BMVDR with the noisy reference microphone signals, enabling to control the binaural cues of the background noise.
Aiming at merging the advantages of both extensions, in this paper we propose the BLCMV with partial noise estimation (BLCMV-N).
We show that the output signals of the BLCMV-N can be interpreted as a mixture between the noisy reference microphone signals and the output signals of a BLCMV using an adjusted interference scaling parameter.
We provide a theoretical comparison between the BMVDR, the BLCMV, the BMVDR-N and the proposed BLCMV-N in terms of noise and interference reduction performance and binaural cue preservation.
Experimental results using recorded signals as well as the results of a perceptual listening test show that the BLCMV-N is able to preserve the binaural cues of an interfering source (like the BLCMV), while enabling to trade off between noise reduction performance and binaural cue preservation of the background noise (like the BMVDR-N).
\end{abstract}


%
\IEEEpeerreviewmaketitle

\section{Introduction}
%
%
%
%
\IEEEPARstart{B}{eamforming} algorithms for head-mounted assistive hearing devices (e.g., hearing aids, earbuds and hearables) are crucial to improve speech quality and speech intelligibility in noisy acoustic environments.
Assuming a binaural configuration where both devices exchange their microphone signals, the information captured by all microphones on both sides of the head can be exploited \cite{Hamacher2008,Doclo2015,Doclo2018}.
Besides reducing interfering sources (e.g., competing speakers) and background noise (e.g., diffuse babble noise), another important objective of a binaural beamforming algorithm is the preservation of the listener's spatial impression of the acoustic scene.
This can be achieved by preserving the binaural cues of all sound sources, i.e., the interaural level difference (ILD) and the interaural time difference (ITD) for coherent sources (desired source and interfering sources) and the interaural coherence (IC) for incoherent sound fields (background noise) \cite{Blauert:1997}.
Binaural cues play a major role for spatial perception, i.e., to localize sound sources and to determine the spatial width or diffuseness of a sound field \cite{Kurozumi:1983}, and are very important for speech intelligibility due to so-called binaural unmasking \cite{Bronkhorst1988,Hawley2004}.

Unlike monaural beamforming algorithms, binaural beamforming algorithms need to generate two output signals (i.e., one for each ear), hence typically processing all available microphone signals from both devices by two different spatial filters \cite{Welker1997,Aichner2007,Klasen2007,Cornelis2010,Hadad2015:2,Marquardt2015:2,Hadad2016,Koutrouvelis2017,Marquardt2018,Goessling2018_iwaenc_b,Asad2019,Corey2020}.
A frequently used binaural beamforming algorithm is the binaural minimum variance distortionless response (BMVDR) beamformer, which aims at minimizing the power spectral density (PSD) of the noise component in the output signals while preserving the desired source component in the reference microphone signals on the left and the right device \cite{Doclo2015,Doclo2018,Cornelis2010}.
While the BMVDR provides a good noise reduction performance and preserves the binaural cues of the desired source, it does not allow to control the reduction of the interfering sources and distorts the binaural cues of the undesired sources (interfering sources and background noise).
More specifically, after applying the BMVDR the binaural cues of the undesired sources are equal to the binaural cues of the desired source, such that all sources are perceived as coming from the same direction, which is obviously undesired.
Hence, several extensions of the BMVDR have been proposed.
On the one hand, the binaural linearly constrained minimum variance (BLCMV) beamformer uses additional interference reduction constraints, enabling to control the reduction of the interfering sources while preserving the binaural cues of the interfering sources in addition to the desired source by means of interference scaling parameters \cite{Hadad2015:2,Hadad2016,Hadad2016comparison,Goessling2018_iwaenc_b}.
However, due to the additional constraints there are less degrees of freedom available for noise reduction, such that the noise reduction performance for the BLCMV is lower than for the BMVDR.
Furthermore, it is not possible to explicitly trade off between noise reduction performance and binaural cue preservation of the background noise.
On the other hand, the BMVDR with partial noise estimation (BMVDR-N) aims for the noise component in the output signals to be equal to a scaled version of the noise component in the reference microphone signals while preserving the desired source component in the reference microphone signals \cite{Doclo2018,Klasen2007,Cornelis2010,Marquardt2018}.
It has been shown that the output signals of the BMVDR-N can be interpreted as a mixture between the output signals of the BMVDR and the noisy reference microphone signals, i.e., the BMVDR-N provides a trade-off between noise reduction performance and binaural cue preservation of the background noise.
While for (incoherent) background noise the BMVDR-N showed promising results \cite{Marquardt2018,Goessling2020_tih}, the effect of partial noise estimation on a (coherent) interfering source strongly depends on the position of the interfering source relative to the desired source and is harder to control \cite{Cornelis2010}.

Aiming at merging the advantages of the BLCMV and the BMVDR-N, i.e., preserving the binaural cues of the interfering sources and controlling the reduction of the interfering sources as well as the binaural cues of the background noise, in this paper we propose the BLCMV with partial noise estimation (BLCMV-N).
First, we derive two decompositions for the BLCMV-N which reveal differences and similarities between the BLCMV-N and the BLCMV.
We show that the output signals of the BLCMV-N can be interpreted as a mixture between the noisy reference microphone signals and the output signals of a BLCMV using an adjusted interference scaling parameter.
We then analytically derive the performance of the BLCMV-N in terms of noise and interference reduction performance and binaural cue preservation.
We show that the output signal-to-noise ratio (SNR) of the BLCMV-N is smaller than or equal to the output SNR of the BLCMV and derive the optimal interference scaling parameter maximizing the output SNR of the BLCMV-N.
The derived analytical expressions are first validated using measured anechoic acoustic transfer functions (ATFs).
In addition, more realistic experiments are performed using recorded signals for a binaural hearing device in a reverberant cafeteria with one interfering source and multi-talker babble noise.
Both the objective performance measures as well as the results of a perceptual listening test with 13 normal-hearing participants show that the proposed BLCMV-N is able to preserve the binaural cues and hence the spatial impression of the interfering source (like the BLCMV), while trading off between noise reduction performance and binaural cue preservation of the background noise (like the BMVDR-N).

The remainder of this paper is organized as follows.
In Section~\ref{sec:config} we introduce the considered binaural hearing device configuration and the used objective performance measures.
In Section~\ref{sec:BinauralNoiseReduction} we briefly review several binaural beamforming algorithms, namely the BMVDR, the BLCMV and the BMVDR-N.
In Section~\ref{sec:lcmvn} we present the BLCMV-N and derive two decompositions.
In Section~\ref{sec:performance} we provide a detailed theoretical analysis of the proposed BLCMV-N in terms of noise and interference reduction performance and binaural cue preservation.
In Section \ref{sec:simulations} we first validate the analytical expressions using anechoic ATFs, followed by simulations and a perceptual listening test using realistic recordings in a reverberant room.
\section{Hearing Device Configuration}
\label{sec:config}
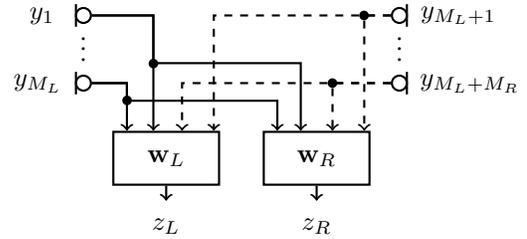
\begin{figure}[t]
    \centering
    \begin{tikzpicture}[auto, thick, node distance=0.9cm]
  \tikzset{
    block/.style = {draw, thick, rectangle, minimum height = 2em, minimum width = 4em},
    mic/.style = {draw, circle, inner sep=2pt},
    mem/.style = {draw, line}
  }

  \newcommand{\lenL}{-1.2}
  \newcommand{\lenR}{3.2}
  \newcommand{\lenM}{0.3}

  \draw
  node at (0,0)[block, name=wL]{$\wL$}
  node at (2,0)[block, name=wR]{$\wR$}
  node at (-1.1,1)[mic] (mic1){}
  node [mic, above of=mic1] (mic2){}
  node at (3.1,1)[mic] (mic3){}
  node [mic, above of=mic3] (mic4){};

  \draw (\lenL,0.85)--(\lenL,0.85+\lenM);
  \draw (\lenL,1.75)--(\lenL,1.75+\lenM);
  \draw (\lenR,0.85)--(\lenR,0.85+\lenM);
  \draw (\lenR,1.75)--(\lenR,1.75+\lenM);

  \coordinate (C1) at (wL.145);
  \coordinate (C2) at (wL.115);
  \coordinate (C3) at (wL.60);

  \draw[->] (mic1) -| (C1);
  \draw[->] (mic2) -| (C2);
  \draw[->] (mic1) -| ($(C1) + (0,0.4)$) -| (wR.145);
  \draw[->] (mic2) -| ($(C2) + (0,0.9)$) -| (wR.120);

  \draw[->,dashed] (mic3) -| (wL.60);
  \draw[->,dashed] (mic4) -| (wL.30);
  \draw[->,dashed] (mic3) -| (wR.60);
  \draw[->,dashed] (mic4) -| (wR.30);

  \draw[->] (wL.south) -| +(0, -0.2cm);
  \draw[->] (wR.south) -| +(0, -0.2cm);

  \draw node [above=0.05cm of mic1] {$\vdots$};
  \draw node [above=0.05cm of mic3] {$\vdots$};

  \draw node [left=0.14cm of mic2] {$y_{1}$};
  \draw node [right=0.03cm of mic3] {$y_{\ML+\MR}$};

  \draw node [right=0.04cm of mic4] {$y_{\ML + 1}$};
  \draw node [left=0.04cm of mic1] {$y_{\ML}$};

  \draw node [below of=wL] {$z_{L}$};
  \draw node [below of=wR] {$z_{R}$};

  \draw node [draw, circle, inner sep=1pt, fill=black] at (-0.52,0.77) {};
  \draw node [draw, circle, inner sep=1pt, fill=black] at (-0.17,1.26) {};
  \draw node [draw, circle, inner sep=1pt, fill=black] at (2.63,1.9) {};
  \draw node [draw, circle, inner sep=1pt, fill=black] at (2.21,1) {};

\end{tikzpicture}
    \vspace{-3mm}
    \caption{Binaural hearing device configuration with $\ML$ microphones on the left side and $\MR$ microphones the right side.}
    \label{fig:configuration}
\end{figure}
In Section~\ref{sec:signalmodel} the considered binaural hearing device configuration and the signal model are introduced.
In Sections~\ref{sec:psds} and \ref{sec:binauralcues} the objective performance measures and the binaural cues are defined.
\subsection{Signal Model}
\label{sec:signalmodel}
Consider the binaural hearing device configuration depicted in Figure \ref{fig:configuration} with $\ML$ microphones on the left side and $\MR$ microphones on the right side, i.e., $M = \ML + \MR$ microphones in total.
In this paper we consider an acoustic scenario with one desired source (target speaker) and one interfering source (competing speaker) in a noisy and reverberant environment, where the background noise is assumed to be incoherent (e.g., diffuse babble noise, sensor noise).

In the frequency-domain, the $m$-th microphone signal $y_m(\omega)$ can be decomposed as
\begin{equation}
	\label{eq:ym}
    y_m(\omega) = x_m(\omega) + u_m(\omega) + n_m(\omega) = x_m(\omega) + v_m(\omega)\, ,
\end{equation}
with $\omega$ the normalized (radian) frequency, $x_m(\omega)$ the desired source component, $u_m(\omega)$ the interfering source component and $n_m(\omega)$ the noise component in the $m$-th microphone signal.
The undesired component $v_m(\omega)$ is defined as the sum of the interfering source component $u_m(\omega)$ and the noise component $n_m(\omega)$.
For the sake of conciseness, we omit the variable $\omega$ in the remainder of the paper wherever possible.
The $M$-dimensional noisy input vector containing all microphone signals is defined as
\begin{equation} \label{eq:vecy}
    \vecy = \left[ y_1,\,\dots,\,y_{\ML},\,y_{\ML+1},\,\dots,\,y_M \right]^T\,,
\end{equation}
where $(\cdot)^T$ denotes the transpose.
Using \eqref{eq:ym}, this vector can be written as
\begin{equation}
	\label{eq:vecyDecomp}
	\vecy = \vecx + \vecu + \vecn = \vecx + \vecv \, ,
\end{equation}
where $\vecx$, $\vecu$, $\vecn$ and $\vecv$ are defined similarly as $\vecy$ in \eqref{eq:vecy}.

For the considered acoustic scenario, the desired source component and the interfering source component can be written as
\begin{equation}
	\label{eq:atfSchreibweise}
	\vecx = \Sx\veca\, , \quad \vecu = \Su\vecb \, ,
\end{equation}
where $\Sx$ and $\Su$ denote the desired source signal and the interfering source signal, respectively, and $\veca$ and $\vecb$ denote $M$-dimensional ATF vectors, containing the ATFs between the microphones and the desired source and the interfering source, respectively.
It should be noted that the ATFs include reverberation, microphone characteristics and the head-shadow effect.

Without loss of generality, the first microphone on each side is defined as the so-called reference microphone.
To simplify the notation, the reference microphone signals $y_1$ and $y_{\ML+1}$ are denoted as $\YL$ and $\YR$, i.e.,
\begin{equation}
	\label{eq:refmics}
	\YL = \veceL^T\vecy \, , \quad \YR = \veceR^T\vecy \, ,
\end{equation}
where $\veceL$ and $\veceR$ denote $M$-dimensional selection vectors with all elements equal to 0 except one element equal to 1, i.e., $\veceL(1) = 1$ and $\veceR(\ML+1) = 1$.
Using \eqref{eq:vecyDecomp}, \eqref{eq:atfSchreibweise} and \eqref{eq:refmics}, the reference microphone signals can be written as
\begin{IEEEeqnarray}{rCl}
	\label{eq: y_L}
	y_L &=& x_L + \underbrace{u_L + n_L}_{v_L} = a_L s_x + b_L s_u + n_L \, , \\[5pt]
	\label{eq: y_R}
	y_R &=& x_R + \underbrace{u_R + n_R}_{v_R} = a_R s_x + b_R s_u + n_R \, .
\end{IEEEeqnarray}
%
The noisy input covariance matrix $\matRy$, the desired source covariance matrix $\matRx$, the interfering source covariance matrix $\matRu$ and the noise covariance matrix $\matRn$ are defined as
\begin{align}
	\label{eq:Ry}
	\matRy &= \mathcal{E}\{ \vecy\vecy^H \} \, , \;	\matRx = \mathcal{E}\{ \vecx\vecx^H \} \, ,\\
	\label{eq:Ru}
	\matRu &= \mathcal{E}\{ \vecu\vecu^H \} \, , \;	\matRn = \mathcal{E}\{\vecn\vecn^H\} \, ,
\end{align}
with $\mathcal{E}\{\cdot\}$ the expected value operator and $(\cdot)^H$ the conjugate transpose.
Assuming statistical independence between all signal components, $\matRy$ can be written as
\begin{equation}
	\matRy = \matRx + \matRu + \matRn = \matRx + \matRv \, ,
\end{equation}
with $\matRv$ the undesired covariance matrix.
Using \eqref{eq:atfSchreibweise}, \eqref{eq:Ry} and \eqref{eq:Ru}, the desired source covariance matrix and the interfering source covariance matrix can be written as rank-1 matrices, i.e.,
\begin{IEEEeqnarray}{C}
	\label{eq:rank1assum}
	\matRx = p_x \mathbf{a}\mathbf{a}^H \,, \quad \matRu = p_u \mathbf{b}\mathbf{b}^H \,,
\end{IEEEeqnarray}
with $p_x = \mathcal{E}\{ \vert s_x \vert^2 \}$ the PSD of the desired source and $p_u = \mathcal{E}\{ \vert s_u \vert^2 \}$ the PSD of the interfering source.
The noise covariance matrix $\matRn$ is assumed to be full-rank, i.e., invertible and positive definite.

The left and the right output signals $\ZL$ and $\ZR$ are obtained by filtering and summing all microphone signals using the $M$-dimensional filter vectors $\wL$ and $\wR$ (cf. Figure~\ref{fig:configuration}), i.e.,
\begin{equation}
	\ZL = \wL^H\vecy\, , \quad \ZR = \wR^H\vecy\, .
\end{equation}
\subsection{Objective Performance Measures}
\label{sec:psds}
The PSD and the cross power spectral density (CPSD) of the desired source component in the left and the right reference microphone signal are given by
\begin{align}
	\label{eq: PSDin}
	p_{x,L}^{\mathrm{in}} &= \mathcal{E}\{ \vert x_L \vert^2 \} = \veceL^T\matRx\veceL = p_x |a\links|^2\, , \\
	p_{x,R}^{\mathrm{in}} &= \mathcal{E}\{ \vert x_R \vert^2 \} = \veceR^T\matRx\veceR = p_x |a\rechts|^2\, , \\
	p_{x,LR}^{\mathrm{in}} &= \mathcal{E}\{ x_L x_R^* \} = \veceL^T\matRx\veceR = p_x a_L a_R^* \, .
\end{align}
Similarly, the output PSD of the desired source component in the left and the right output signal is given by
\begin{equation}
	\label{eq:psdout}
	p_{x,L}^\mathrm{out} = \wL^H\matRx\wL \, , \quad p_{x,R}^\mathrm{out} = \wR^H\matRx\wR \, .
\end{equation}
The same definitions can be applied for the noisy input signal, the interfering source component and the noise component by substituting $\matRx$ with $\matRy$, $\matRu$ or $\matRn$.

The narrowband input SNR in the left and the right reference microphone signal is defined as the ratio of the input PSD of the desired source and noise components, i.e.,
\begin{equation}
	\label{eq: SNRin}
	\SNRinL = \frac{p^{\mathrm{in}}_{x,L}}{p^{\mathrm{in}}_{n,L}} \, , \quad \SNRinR = \frac{p^{\mathrm{in}}_{x,R}}{p^{\mathrm{in}}_{n,R}} \, .
\end{equation}
Similarly, the narrowband output SNR in the left and the right output signal is defined as the ratio of the output PSD of the desired source and noise components, i.e.,
\begin{equation}
	\label{eq:SNRout}
	\SNRoutL = \frac{p_{x,L}^\mathrm{out}}{p_{n,L}^\mathrm{out}} \, , \quad \SNRoutR = \frac{p_{x,R}^\mathrm{out}}{p_{n,R}^\mathrm{out}} \, .
\end{equation}
The SNR improvement (in dB) is defined as $\Delta\mathrm{SNR}_{L/R} = 10 \log_{10} \mathrm{SNR}_{L/R}^\mathrm{out} - 10 \log_{10} \mathrm{SNR}_{L/R}^\mathrm{in}$.

The narrowband input signal-to-interference ratio (SIR) in the left and the right reference microphone signal is defined as the ratio of the input PSD of the desired source and interfering source components, i.e.,
\begin{equation}
	\label{eq: SIRin}
	\SIRinL = \frac{p^{\mathrm{in}}_{x,L}}{p^{\mathrm{in}}_{u,L}} \, , \quad \SIRinR = \frac{p^{\mathrm{in}}_{x,R}}{p^{\mathrm{in}}_{u,R}} \, .
\end{equation}
Similarly, the narrowband output SIR in the left and the right output signal is defined as the ratio of the output PSD of the desired source and interfering source components, i.e.,
\begin{equation} \label{eq:SIRout}
	\SIRoutL = \frac{p_{x,L}^\mathrm{out}}{p_{u,L}^\mathrm{out}} \, , \quad \SIRoutR = \frac{p_{x,R}^\mathrm{out}}{p_{u,R}^\mathrm{out}} \, .
\end{equation}
The SIR improvement (in dB) is defined as $\Delta\mathrm{SIR}_{L/R} = 10 \log_{10} \mathrm{SIR}_{L/R}^\mathrm{out} - 10 \log_{10} \mathrm{SIR}_{L/R}^\mathrm{in}$.
%
\subsection{Binaural Cues}
\label{sec:binauralcues}
For coherent sources (desired source and interfering source) the main binaural cues used by the auditory system are the ILD and the ITD \cite{Blauert:1997}, which can be computed from the so-called interaural transfer function (ITF).
Using \eqref{eq:rank1assum}, the input ITFs of the desired source and the interfering source are given by \cite{Cornelis2010}
\begin{IEEEeqnarray}{rCl}
	\label{eq: inITFx}
	\mathrm{ITF}_x^{\mathrm{in}} &=& \frac{\mathcal{E}\{|x_L|^2\}}{\mathcal{E}\{x_R x_L^*\}} = \frac{\mathbf{e}_L^T \mathbf{R}_x \mathbf{e}_L}{\mathbf{e}_R^T \mathbf{R}_x \mathbf{e}_L} = \frac{a_L}{a_R} \, , \\[5pt]
	\label{eq: inITFu}
	\mathrm{ITF}_u^{\mathrm{in}} &=& \frac{\mathcal{E}\{|u_L|^2\}}{\mathcal{E}\{u_R u_L^*\}} = \frac{\mathbf{e}_L^T \mathbf{R}_u \mathbf{e}_L}{\mathbf{e}_R^T \mathbf{R}_u \mathbf{e}_L} = \frac{b_L}{b_R} \, .
\end{IEEEeqnarray}
Similarly, the output ITFs of the desired source and the interfering source are given by
\begin{IEEEeqnarray}{rCl}
	\label{eq: outITFx}
	\mathrm{ITF}_x^{\mathrm{out}} &=& \frac{\mathbf{w}_L^H \mathbf{R}_x \mathbf{w}_L}{\mathbf{w}_R^H \mathbf{R}_x \mathbf{w}_L} = \frac{\mathbf{w}_L^H \mathbf{a}}{\mathbf{w}_R^H \mathbf{a}} \, , \\[5pt]
	\label{eq: outITFu}
	\mathrm{ITF}_u^{\mathrm{out}} &=& \frac{\mathbf{w}_L^H \mathbf{R}_u \mathbf{w}_L}{\mathbf{w}_R^H \mathbf{R}_u \mathbf{w}_L} = \frac{\mathbf{w}_L^H \mathbf{b}}{\mathbf{w}_R^H \mathbf{b}}.
\end{IEEEeqnarray}
The ILD and the ITD can be calculated from the ITF as \cite{Cornelis2010}
\begin{IEEEeqnarray}{C}
	\label{eq: ILDandITD}
	\mathrm{ILD} = \vert \mathrm{ITF} \vert^2  \, , \quad	\mathrm{ITD} = \frac{\angle\mathrm{ITF}}{\omega} \, ,
\end{IEEEeqnarray}
with $\angle(\cdot)$ denoting the unwrapped phase.

For an incoherent sound field (background noise), ILD and ITD cues are not very descriptive, but the IC is known to play a major role for spatial perception (e.g., spatial width or diffuseness) \cite{Blauert:1997}.
The input IC of the noise component is defined as
\begin{equation}
	\label{eq: inICn}
	\mathrm{IC}_n^\mathrm{in} = \frac{\mathcal{E}\{ n_L n_R^* \}}{\sqrt{\mathcal{E}\{ |n_L|^2 \}} \sqrt{\mathcal{E}\{ |n_R|^2 \}}} = \frac{\mathbf{e}_L^T \mathbf{R}_n \mathbf{e}_R}{\sqrt{\mathbf{e}_L^T\mathbf{R}_n\mathbf{e}_L} \sqrt{\mathbf{e}_R^T\mathbf{R}_n\mathbf{e}_R}} \, , \nonumber
\end{equation}
while the output IC of the noise component is defined as
\begin{equation}
	\label{eq: outICn}
	\mathrm{IC}_n^\mathrm{out} = \frac{\mathbf{w}_L^H \mathbf{R}_n \mathbf{w}_R}{\sqrt{ \mathbf{w}_L^H\mathbf{R}_n\mathbf{w}_L } \sqrt{ \mathbf{w}_R^H\mathbf{R}_n\mathbf{w}_R }} \, .
\end{equation}
Because the IC is typically complex-valued, the magnitude-squared coherence (MSC) is often used.
The input and the output MSC of the noise component are defined as
\begin{equation} \label{eq:MSCout}
	\mathrm{MSC}^{\mathrm{in}}_n = |\mathrm{IC}_n^{\mathrm{in}}|^2 \,, \quad \mathrm{MSC}_n^{\mathrm{out}} = |\mathrm{IC}_n^{\mathrm{out}}|^2 \, .
\end{equation}
An MSC of 1 corresponds to a coherent source perceived as a distinct point source, while smaller MSC values correspond to a broader or even diffuse sound field impression \cite{Blauert:1997}.
%
%
%
\section{Binaural Beamforming Algorithms}
\label{sec:BinauralNoiseReduction}
In this section we briefly review three state-of-the-art binaural beamforming algorithms, namely the BMVDR beamformer, the BLCMV beamformer and the BMVDR-N beamformer.
We discuss the performance of these beamforming algorithms in terms of noise and interference reduction performance and binaural cue preservation.
For the sake of conciseness, we only show expressions for the left hearing device, denoted by the subscript ${L}$.
It should be noted that all expressions can also be formulated for the right hearing device by changing the subscript to ${R}$.
%
\subsection{BMVDR Beamformer}
\label{sec:mvdr}
The BMVDR aims at minimizing the output PSD of the noise component while preserving the desired source component in the reference microphone signals \cite{Doclo2015,Doclo2018,Cornelis2010}.
The constrained optimization problem for the left filter vector is given by
\begin{equation}
	\label{eq: cost_BMVDR_L}
	\boxed{
	\min_{\mathbf{w}_L} \, \mathcal{E}\{ \vert \mathbf{w}_L^H\mathbf{n} \vert^2 \} \quad \text{s.t.} \quad \mathbf{w}_L^H \mathbf{x} = x_L}
\end{equation}
Using \eqref{eq:atfSchreibweise}, \eqref{eq: y_L} and \eqref{eq:Ru}, the solution of \eqref{eq: cost_BMVDR_L} is equal to \cite{Doclo2015,VanVeen1988a,Gannot2017}
\begin{equation}
	\label{eq: bmvdr_L}
	\boxed{
	\mathbf{w}_{\mathrm{BMVDR},L} = \frac{\mathbf{R}_n^{-1}\mathbf{a}}{\gamma_a} a_L^*}
\end{equation}
with
\begin{equation}
	\label{eq: gamma_a}
	\gamma_a = \mathbf{a}^H \mathbf{R}_n^{-1} \mathbf{a} \, .
\end{equation}
It should be noted that the BMVDR can also be defined using the undesired covariance matrix $\matRv$ instead of the noise covariance matrix $\matRn$.
However, since $\matRv$ is considerably more difficult to estimate or model in practice than $\matRn$, in this paper we only consider the BMVDR using $\matRn$ in \eqref{eq: bmvdr_L}.

%
By substituting \eqref{eq: bmvdr_L} in \eqref{eq:SNRout} and \eqref{eq:SIRout}, it has been shown in \cite{Doclo2018,Cornelis2010} that the output SNR and the output SIR of the BMVDR are equal to
\begin{IEEEeqnarray}{rCl}
	\label{eq: outSNR_bmvdr}
	\mathrm{SNR}_{\mathrm{BMVDR},L}^\mathrm{out} &=& p_{x} \gamma_a \, , \\
	\label{eq: sir_bmvdr}
	\mathrm{SIR}_{\mathrm{BMVDR},L}^\mathrm{out} &=& \frac{p_{x}}{p_{u}} \frac{\vert \gamma_a \vert^2}{\vert \gamma_{ab} \vert^2} \, ,
\end{IEEEeqnarray}
with $\gamma_a$ defined in \eqref{eq: gamma_a} and
\begin{equation}
	\label{eq: gamma_ab}
	\gamma_{ab} = \mathbf{a}^H \mathbf{R}_n^{-1} \mathbf{b} \, .
\end{equation}
Although the BMVDR yields the largest output SNR among all distortionless binaural beamforming algorithms, the output SIR depends on the relative position of the interfering source to the desired source, cf. \eqref{eq: gamma_ab}.

%
As shown in \cite{Doclo2018,Cornelis2010,Marquardt2015:2}, the BMVDR preserves the binaural cues of the desired source, i.e.,
\begin{IEEEeqnarray}{C}
	\mathrm{ITF}_{\mathrm{BMVDR},x}^\mathrm{out} = \frac{a_L}{a_R} = \mathrm{ITF}_x^\mathrm{in} \, ,
\end{IEEEeqnarray}
but distorts the binaural cues of the undesired sources, i.e., for the interfering source
\begin{equation}
	\mathrm{ITF}_{\mathrm{BMVDR},u}^\mathrm{out} = \frac{a_L}{a_R} = \mathrm{ITF}_x^\mathrm{in} \, ,
\end{equation}
and for the background noise
\begin{equation}
	\label{eq: ICoutBMVDR}
	\mathrm{IC}_{\mathrm{BMVDR},n}^\mathrm{out} = \mathrm{IC}_{x}^\mathrm{in} = e^{j \, a_L/a_R} \, , \; \mathrm{MSC}_{\mathrm{BMVDR},n}^\mathrm{out} = 1 \, .
\end{equation}
Hence, at the output of the BMVDR the interfering source and the (incoherent) background noise are perceived as coming from the direction of the desired source, which is obviously undesired in terms of spatial awareness.
%
%
%
\subsection{BLCMV Beamformer}
\label{sec:lcmv}
In addition to preserving the desired source component in the reference microphone signals, the BLCMV preserves a scaled version of the interfering source component in the reference microphone signals	 while minimizing the output PSD of the noise component \cite{Hadad2015:2,Hadad2016}.
The constrained optimization problem for the left filter vector is given by \cite{Hadad2016}
\begin{equation}
	\label{eq: cost_blcmv_L}
	\boxed{
	\min_{\mathbf{w}_L} \, \mathcal{E}\{ \vert \mathbf{w}_L^H\mathbf{n} \vert^2 \} \quad \text{s.t.} \quad \mathbf{w}_L^H\mathbf{x} = x_L \, , \; \mathbf{w}_L^H\mathbf{u} = \delta u_L } 
\end{equation}
with $0 < \delta \leq 1$ the (real-valued) interference scaling parameter.
Using \eqref{eq:atfSchreibweise}, \eqref{eq: y_L} and \eqref{eq:Ru}, the solution of \eqref{eq: cost_blcmv_L} is equal to \cite{Hadad2016}
\begin{equation}
	\label{eq: blcmv_L}
	\boxed{
	\mathbf{w}_{\mathrm{BLCMV},L} = \mathbf{R}_n^{-1}\mathbf{C}\left(\mathbf{C}^H\mathbf{R}_n^{-1}\mathbf{C}\right)^{-1}\mathbf{g}_L}
\end{equation}
with the constraint matrix $\matC$ and the left response vector $\mathbf{g}_L$ defined as
\begin{equation}
	\label{eq: CandgL}
	\matC = \left[\veca , \; \vecb \right] \, , \quad \mathbf{g}_L = \left[a_L , \; \delta b_L \right]^H \, .
\end{equation}

%
By substituting \eqref{eq: blcmv_L} in \eqref{eq:SNRout}, it has been shown in \cite{Hadad2016} that the output SNR of the BLCMV is equal to
\begin{equation}
	\label{eq: outSNR_blcmv_L}
	\mathrm{SNR}_{\mathrm{BLCMV},L}^\mathrm{out} = \frac{p_x |a\links|^2}{\mathbf{e}_L^T \mathbf{R}_{xu,1} \mathbf{e}_L} \, ,
\end{equation}
with
\begin{IEEEeqnarray}{C}
	\label{eq: Rxu1}
	\mathbf{R}_{xu,1} = \frac{1}{1- \Psi} \left[ \frac{\mathbf{a}\mathbf{a}^H}{\gamma_a} + \delta^2 \frac{\mathbf{b}\mathbf{b}^H}{\gamma_b} - 2\Psi\delta\Re\left\{\frac{\mathbf{a}\mathbf{b}^H}{\gamma_{ab}^*}\right\}\right] \, , \IEEEeqnarraynumspace \\
	\label{eq: gamma_b_Psi}
	\gamma_b = \mathbf{b}^H \mathbf{R}_n^{-1} \mathbf{b} \, , \quad \Psi = \frac{\vert \gamma_{ab} \vert^2}{\gamma_a \gamma_b} \, ,
\end{IEEEeqnarray}
where $\Re \{\cdot\}$ denotes the real part of a complex number.
The output SNR of the BLCMV in \eqref{eq: outSNR_blcmv_L} is smaller than or equal to the output SNR of the BMVDR in \eqref{eq: outSNR_bmvdr}, since less degrees of freedom are available for noise reduction.
In addition, the output SIR of the BLCMV is equal to \cite{Hadad2016}
\begin{equation}
	\label{eq:lcmvSIRout}
	\SIRoutLlcmv = \frac{1}{\delta^2} \SIRinL \, ,
\end{equation}
which can hence be directly controlled by the interference scaling parameter $\delta$.

%
As shown in \cite{Hadad2016}, the BLCMV preserves the binaural cues of both the desired source and the interfering source, i.e.,
\begin{IEEEeqnarray}{rClCl}
	\mathrm{ITF}_{\mathrm{BLCMV},x}^{\mathrm{out}} &=& \frac{a\links}{a\rechts} &=& \mathrm{ITF}_{{x}}^{\mathrm{in}} \, , \\
	\mathrm{ITF}_{\mathrm{BLCMV},u}^{\mathrm{out}} &=& \frac{b\links}{b\rechts} &=& \mathrm{ITF}_{{u}}^{\mathrm{in}}
\end{IEEEeqnarray}
and the output MSC of the noise component is equal to
\begin{equation}
	\label{eq: ICn_blcmv}
	\mathrm{MSC}_{\mathrm{BLCMV},n}^\mathrm{out} = \left\vert \frac{\mathbf{e}_L^T \mathbf{R}_{xu,1} \mathbf{e}_R}{\sqrt{\mathbf{e}_L^T \mathbf{R}_{xu,1} \mathbf{e}_L} \sqrt{\mathbf{e}_R^T \mathbf{R}_{xu,1} \mathbf{e}_R } } \right\vert^2 \, .
\end{equation}
Because $\mathbf{R}_{xu,1}$ in \eqref{eq: Rxu1} is a rank-2 matrix, it has been shown in \cite{Hadad2016} that the output MSC of the noise component is smaller than 1 but is not equal to the input MSC of the noise component.
Furthermore, it should be noted that the output MSC of the noise component depends on the relative position of the interfering source to the desired source, cf. \eqref{eq: Rxu1} and \eqref{eq: gamma_b_Psi}, such that it is not straightforward to control the binaural cues of the background noise.

%
%
\subsection{BMVDR-N beamformer}
\label{sec:mvdrn}
In addition to preserving the desired source component in the reference microphone signals, the BMVDR with partial noise estimation (BMVDR-N) aims at preserving a scaled version of the noise component in the reference microphone signals \cite{Doclo2018,Cornelis2010,Klasen2007}.
The constrained optimization problem for the left filter vector is given by 
\begin{equation}
	\label{eq: cost_bmvdrn_L}
	\boxed{
	\min_{\mathbf{w}_L} \, \mathcal{E}\left\{ \left\vert\mathbf{w}_L^H\mathbf{n} - \eta n_L \right\vert^2 \right\} \quad \text{s.t.} \quad \mathbf{w}_L^H \mathbf{x} = x_L }
\end{equation}
with $0 \leq \eta \leq 1$ the (real-valued) mixing parameter.
It has been shown in \cite{Cornelis2010} that the solution of \eqref{eq: cost_bmvdrn_L} is equal to
\begin{equation}
	\label{eq: bmvdrn_L}
	\boxed{
	\mathbf{w}_{\mathrm{BMVDR-N},L} = (1-\eta) \mathbf{w}_{\mathrm{BMVDR},L} + \eta \mathbf{e}_L }
\end{equation}
with $\mathbf{w}_{\mathrm{BMVDR},L}$ defined in \eqref{eq: bmvdr_L}.
Hence, the output signals of the BMVDR-N can be interpreted as a mixture between the noisy reference microphone signals (scaled with $\eta$) and the output signals of the BMVDR (scaled with $1-\eta$).
For $\eta=0$, the BMVDR-N is equal to the BMVDR, whereas for $\eta=1$, no beamforming is applied.

%
Since the output signals of the BMVDR are mixed with the noisy reference microphone signals, the output SNR of the BMVDR-N is always smaller than or equal to the output SNR of the BMVDR \cite{Cornelis2010}, i.e.,
\begin{equation}
	\label{eq: SNRoutMVDRNvsSNRoutMVDR}
	\SNRoutLmvdrn \leq \SNRoutLmvdr
\end{equation}
and decreases with increasing $\eta$.
By substituting \eqref{eq: bmvdrn_L} in \eqref{eq:SIRout}, it can be shown that the output SIR of the BMVDR-N is equal to
\begin{equation}
	\label{eq: outSIR_bmvdrn_L}
	\mathrm{SIR}_{\mathrm{BMVDR-N},L}^\mathrm{out} = \frac{p_{x}}{p_{u}} \frac{\vert a_L \vert^2}{\mathbf{e}_L^T \mathbf{R}_{xu,2} \mathbf{e}_L } \, ,
\end{equation}
with
\begin{equation}
	\label{eq: Rxu2}
	\mathbf{R}_{xu,2} = (1-\eta)^2 \frac{\vert \gamma_{ab} \vert^2}{\vert \gamma_a \vert^2} \mathbf{a}\mathbf{a}^H + \eta^2 \mathbf{b}\mathbf{b}^H + (\eta - \eta^2) 2 \Re\{ \mathbf{a}\mathbf{b}^H \frac{\gamma_{ab}}{\gamma_a} \}\, . \nonumber
\end{equation}

%
As shown in \cite{Cornelis2010,Marquardt2018}, the BMVDR-N preserves the binaural cues of the desired source, i.e.,
\begin{equation}
	\mathrm{ITF}_{\mathrm{BMVDR-N},x}^{\mathrm{out}} = \frac{a\links}{a\rechts} = \mathrm{ITF}_{{x}}^{\mathrm{in}} \, .
\end{equation}
By substituting \eqref{eq: bmvdrn_L} in \eqref{eq: outITFu} and \eqref{eq: outICn}, it has been shown in \cite{Marquardt2018} and \cite{Hadad2016comparison} that the output ITF of the interfering source is equal to
\begin{equation}
	\label{eq: outITFu_bmvdrn}
	\mathrm{ITF}_{\mathrm{BMVDR-N},u}^\mathrm{out} = \frac{(1-\eta) a_L \frac{\gamma_{ab}}{\gamma_a} + \eta b_L}{(1-\eta) a_R \frac{\gamma_{ab}}{\gamma_a} + \eta b_R} \, ,
\end{equation}
and the output MSC of the noise component is equal to
\begin{align}
	\label{eq: outMSC_bmvdrn}
	&\mathrm{MSC}_{\mathrm{BMVDR-N},n}^\mathrm{out} = \\ \nonumber
	&\frac{ \left\vert \frac{1-\eta^2}{p_{x} \gamma_a} p^{\mathrm{in}}_{x,LR} + \eta^2 p^{\mathrm{in}}_{n,LR} \right\vert^2 }{ \left( \frac{1-\eta^2}{p_{x} \gamma_a} p^{\mathrm{in}}_{x,L} + \eta^2 p^{\mathrm{in}}_{n,L} \right) \left( \frac{1-\eta^2}{p_{x} \gamma_a} p^{\mathrm{in}}_{x,R} + \eta^2 p^{\mathrm{in}}_{n,R} \right) } \, .
\end{align}
It can be seen from \eqref{eq: outITFu_bmvdrn} and \eqref{eq: outMSC_bmvdrn} that only for $\eta=1$ the binaural cues of the undesired sources (interfering source and background noise) are preserved, whereas for $\eta=0$ the binaural cues of the undesired sources are equal to the binaural cues of the desired source (as for the BMVDR).
The mixing parameter $\eta$ hence allows to trade off between noise reduction performance and binaural cue preservation of the background noise, or in other words control the binaural cues of the background noise.
Furthermore, it should be noted that the interference reduction performance in \eqref{eq: outSIR_bmvdrn_L} and the output ITF of the interfering source in \eqref{eq: outITFu_bmvdrn} do not only depend on the mixing parameter $\eta$ but also on the relative position of the interfering source to the desired source, such that it is not straightforward to control both.
%
%
%
\section{BLCMV with partial noise estimation}
\label{sec:lcmvn}
Aiming at merging the advantages of the BLCMV and the BMVDR-N, i.e., preserving the binaural cues of the interfering source and controlling the binaural cues of the background noise, in Section~\ref{sec:lcmvnderiv} we present the BLCMV beamformer with partial noise estimation (BLCMV-N).
Similarly as for the BLCMV in \cite{Hadad2016}, in Sections \ref{sec:decomp1} and \ref{sec:decomp2} we derive two decompositions for the BLCMV-N which reveal differences and similarities between the BLCMV-N and the BLCMV.
%
\subsection{BLCMV-N Beamformer}
\label{sec:lcmvnderiv}
\begin{figure}[t]
  \centering
  \includegraphics{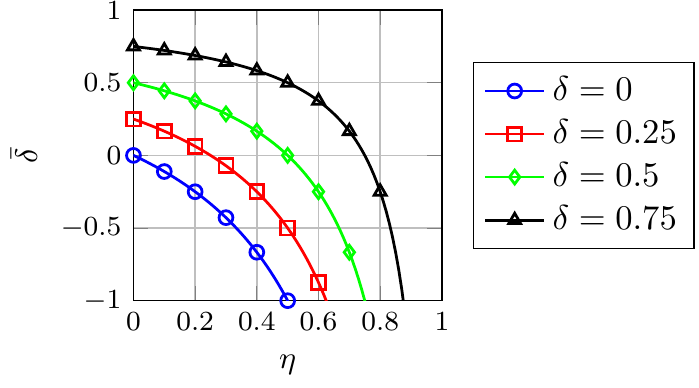}	
  \vspace{-2mm}
  \caption{Adjusted interference scaling parameter $\bar{\delta}$ as a function of $\eta$ for different values of $\delta$.}
  \label{fig:deltaTilde}
\end{figure}

Compared to the BMVDR in \eqref{eq: cost_BMVDR_L}, the BLCMV-N uses an additional constraint to preserve a scaled version of the interfering source component in the reference microphone signals, like the BLCMV in \eqref{eq: cost_blcmv_L}, and aims at preserving a scaled version of the noise component in the reference microphone signals, like the BMVDR-N in \eqref{eq: cost_bmvdrn_L}.
The constrained optimization problem for the left filter vector is given by
\begin{equation} \label{eq:costBLCMVN}
	\boxed{
	\min_{\wL}\mathcal{E}\left\{ \left| \wL^H\vecn - \eta n\links \right|^2\right\} \quad \text{s.t.} \quad \mathbf{w}_L^H\mathbf{x} = x_L \, , \; \mathbf{w}_L^H\mathbf{u} = \delta u_L }
\end{equation}
The solution of \eqref{eq:costBLCMVN} is equal to (see Appendix~\ref{App1})
\begin{equation}
	\label{eq: blcmvn_L}
	\boxed{
	\begin{split}
		\wLblcmvn &= \\
		\eta \veceL + (1&-\eta)\matRn^{-1}\matC\left(\matC^H\matRn^{-1}\matC\right)^{-1} \begin{bmatrix}
			a\links^*\\
			\bar{\delta} b\links^*
		\end{bmatrix}
	\end{split}}
\end{equation}
with $\mathbf{C}$ defined in \eqref{eq: CandgL} and the \textit{adjusted interference scaling parameter} $\bar{\delta}$ equal to
\begin{equation}
	\label{eq: adjDelta}
	\bar{\delta} = \frac{\delta - \eta}{1 - \eta} \, .
\end{equation}
Hence, the output signals of the BLCMV-N can be interpreted as a mixture between the noisy reference microphone signals (scaled with $\eta$) and the output signals of a BLCMV (scaled with $1-\eta$) using the adjusted interference scaling parameter $\bar{\delta}$ in \eqref{eq: adjDelta} instead of the interference scaling parameter $\delta$.
For $\eta = 0$, the BLCMV-N is equal to the BLCMV in \eqref{eq: blcmv_L} with $\bar{\delta} = \delta$, whereas for $\eta = 1$, it should be realized that only if $\delta=1$ no beamforming is applied.
Since mixing with the reference microphone signals not only affects the noise component but also the interfering source component, the adjusted interference scaling parameter $\bar{\delta}$ depends on both the interference scaling parameter $\delta$ as well as the mixing parameter $\eta$ due to the interference reduction constraint in \eqref{eq:costBLCMVN}.
Figure \ref{fig:deltaTilde} depicts $\bar{\delta}$ as a function of $\eta$ for different values of $\delta$.
It can be seen that
\begin{IEEEeqnarray}{C}
	\bar{\delta}(\eta,\delta) = \begin{cases}
		>0\,,& 	\text{for} \; \delta>\eta \\
		<0\,, & 	\text{for} \; \delta<\eta \\
		\phantom{<}\;0\,,	&	\text{for} \; \delta=\eta
	\end{cases} \, .
\end{IEEEeqnarray}
As will be shown in more detail in the following sections, using the parameters $\delta$ and $\eta$ it is possible to control the noise reduction performance, the interference reduction performance and the binaural cues of the background noise for the BLCMV-N.
%
%
\subsection{Decomposition into two BLCMVs}
\label{sec:decomp1}
In \cite{Hadad2016} it has been shown that the BLCMV in \eqref{eq: blcmv_L} can be decomposed as the sum of two sub-BLCMVs, i.e.,
\begin{equation}
	\wLblcmv = \wxl + \delta\wul \, ,
\end{equation}
with
\begin{align}
	\label{eq:sub1}
	\wxl &= \matRn^{-1}\matC \left(\matC^H\matRn^{-1}\matC\right)^{-1}\vecgxl \, ,\\
	\label{eq:sub2}
	\wul &= \matRn^{-1}\matC \left(\matC^H\matRn^{-1}\matC\right)^{-1}\vecgul \, ,
\end{align}
and the respective response vectors
\begin{equation}
	\vecgxl = \begin{bmatrix}
		a\links^*\\
		0
	\end{bmatrix} \, , \quad
	\vecgul = \begin{bmatrix}
		0\\
		b\links^*
	\end{bmatrix} \, .
\end{equation}
The sub-BLCMV $\wxl$ in \eqref{eq:sub1} preserves the desired source component in the reference microphone signals and steers a null towards the interfering source, whereas the sub-BLCMV $\wul$ in \eqref{eq:sub2} preserves the interfering source component in the reference microphone signals and steers a null towards the desired source.
Using \eqref{eq: blcmvn_L}, it can be easily seen that the proposed BLCMV-N can be decomposed as 
\begin{equation}
	\label{eq:decomp1}
	\boxed{
	\wLblcmvn = \eta \veceL + (1 - \eta)\wxl + (\delta - \eta)\wul }
\end{equation}
Hence, the BLCMV-N can be interpreted as a mixture of the reference microphone signals (scaled with $\eta$), a BLCMV that preserves the desired source and rejects the interfering source (scaled with $1-\eta$) and a BLCMV that preserves the interfering source and rejects the desired source (scaled with $\delta - \eta$).
%
Since the scaling of the sub-BLCMV $\wxl$ controls the desired source component without affecting the interfering source component and the scaling of the sub-BLCMV $\wul$ controls the interfering source component without affecting the desired source component \cite{Hadad2016}, it can be directly observed from the scaling factors in \eqref{eq:decomp1} that the desired source component is not distorted and the interfering source component is scaled with $\delta$.
\subsection{Decomposition using Binauralization Postfilters}
\label{sec:decomp2}
In \cite{Hadad2016} it has also been shown that the sub-BLCMV $\wxl$ in \eqref{eq:sub1} for the left hearing device and the sub-BLCMV $\wxr$ for the right hearing device (defined similarly as $\wxl$) can be written using a common spatial filter and two binauralization postfilters as
\begin{equation} \label{eq:wx_1}
	\wxl = \wx a_L^* \, , \quad \wxr = \wx a_R^* \, ,
\end{equation}
with the common desired BLCMV (D-BLCMV) given by
\begin{equation} \label{eq:wx}
	\wx = \frac{1}{1-\Psi} \left(\frac{\matRn^{-1}\veca}{\ga} - \Psi \frac{\matRn^{-1}\vecb}{\gab} \right) \, ,
\end{equation}
and the ATFs $a_L$ and $a_R$ between the desired source and the reference microphones used as binauralization postfilters.
Similarly, the sub-BLCMV $\wul$ in \eqref{eq:sub2} and the sub-BLCMV $\wur$ (defined similarly as $\wul$) can be written as
\begin{equation} \label{eq:wu_1}
	\wul = \wu b_L^* \, , \quad \wur = \wu b_R^* \, ,
\end{equation}
with the common interference BLCMV (I-BLCMV) given by
\begin{equation} \label{eq:wu}
	\wu = \frac{1}{1 - \Psi} \left( \frac{\matRn^{-1}\vecb}{\gb} - \Psi \frac{\matRn^{-1}\veca}{\gab^*} \right) \, ,
\end{equation}
and the ATFs $b_L$ and $b_R$ between the interfering source and the reference microphones used as binauralization postfilters.

\begin{figure}[t]
  \centering
  \begin{tikzpicture}[node distance=1cm, >=latex',every text node part/.style={align=center}, rounded corners=1pt]
    \tikzstyle{block} = [draw, rectangle];
    \tikzstyle{mycirc} = [draw, circle, inner sep=0pt];
    \tikzstyle{dot} = [draw, fill=black, circle, inner sep=1pt];

    \node (input) {$\vecy$};

    \node [dot, right of=input, node distance=0.5cm] (dot1) {};
    \node [dot, right of=dot1, node distance=0.5cm] (dot2) {};

    \node [block, above right of=dot2] (wx) {$\mathbf{w}_x$};
    \node [block, below right of=dot2] (wu) {$\mathbf{w}_u$};
    \node [block, above of=wx] (eL) {$\veceL$};
    \node [block, below of=wu] (eR) {$\veceR$};
    \node [block, right of=wx, node distance=1.5cm, yshift=3.5mm] (binL) {\tiny{$a\links^*$}};
    \node [block, right of=wu, node distance=1.5cm, yshift=3.5mm] (binR) {\tiny{$a\rechts^*$}};
    \node [block, below of=binL, node distance=0.7cm] (binL2) {\tiny{$b\links^*$}};
    \node [block, below of=binR, node distance=0.7cm] (binR2) {\tiny{$b\rechts^*$}};
    \node [mycirc, right of=eR, node distance=2.5cm, label={[label distance=-0.95mm]above:\tiny{$\eta$}}] (mult2) {$\times$};
    \node [mycirc, right of=eL, node distance=2.5cm, label={[label distance=-0.95mm]above:\tiny{$\eta$}}] (mult1) {$\times$};
    \node [mycirc, label={[label distance=-0.95mm]above:\tiny{$1-\eta$}}] (mult3) at (binL-|mult1) {$\times$};
    \node [mycirc, label={[label distance=-0.95mm]above:\tiny{$\delta-\eta$}}] (mult4) at (binL2-|mult1) {$\times$};
    \node [mycirc, label={[label distance=-0.95mm]above:\tiny{$\delta-\eta$}}] (mult5) at (binR2-|mult1) {$\times$};
    \node [mycirc, label={[label distance=-0.95mm]above:\tiny{$1-\eta$}}] (mult6) at (binR-|mult1) {$\times$};

    \node [dot, right of=wu, node distance = 0.75cm] (dot3) {};
    \node [dot, right of=wx, node distance=0.55cm] (dot4) {};

    \node [mycirc, right of=binL, node distance=1.75cm] (sumL) {$+$};
    \node [mycirc, right of=binR2, node distance=1.75cm] (sumR) {$+$};

    \node [right of=sumL] (ZL) {$z_{L}$};
    \node [right of=sumR] (ZR) {$z_{R}$};

    \draw[-] (input)--(dot1)--(dot2);
    \draw[->] (dot1)|-(eL);
    \draw[->] (dot1)|-(eR);
    \draw[->] (dot2)|-(wx);
    \draw[->] (dot2)|-(wu);
    \draw[->] (eR)--(mult2);
    \draw[->] (eL)--(mult1);
    \draw[-] (wu)--(dot3);
    \draw[-] (wx)--(dot4);
    \draw[->] (dot4)|-(binL);    
    \draw[->] (dot4)|-(binR);    
    \draw[->] (dot3)|-(binR2);    
    \draw[->] (dot3)|-(binL2);    
    \draw[->] (binL)--(mult3);
    \draw[->] (mult3)--(sumL);
    \draw[->] (binL2)--(mult4);
    \draw[->] (mult4)-|(sumL);
    \draw[->] (binR)--(mult6);
    \draw[->] (mult6)-|(sumR);
    \draw[->] (binR2)--(mult5);
    \draw[->] (mult5)--(sumR);
    \draw[->] (mult1)-|(sumL);
    \draw[->] (mult2)-|(sumR);
    \draw[->] (sumL)--(ZL);
    \draw[->] (sumR)--(ZR);
    
  \end{tikzpicture}
  \vspace{0mm}
  \caption{Decomposition of the BLCMV-N into a mixture of the reference microphone signals and two BLCMVs with binauralization postfilters.}
  \label{fig:decomp2}
\end{figure}
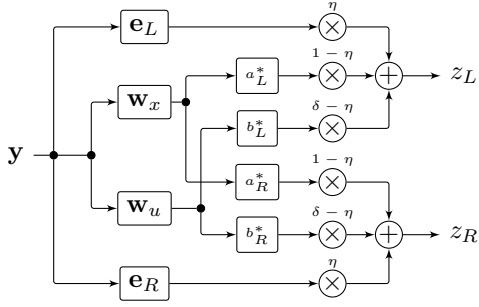

Using \eqref{eq:wx_1} and \eqref{eq:wu_1} in \eqref{eq:decomp1}, the BLCMV-N can be decomposed as
\begin{empheq}[box=\fbox]{align}
	\label{eq:decomp2}
	\wLblcmvn & = \eta \veceL + (1-\eta)a_L^*\wx \\ \nonumber
			& \quad + (\delta - \eta) b_L^* \wu \\
	\label{eq:decomp2R}
	\wRblcmvn & = \eta \veceR + (1-\eta)a_R^*\wx \\ \nonumber
			& \quad + (\delta - \eta) b_R^* \wu
\end{empheq}
Figure \ref{fig:decomp2} depicts this decomposition of the BLCMV-N using common spatial filters and binauralization postfilters.
The output signals of the BLCMV-N can hence be interpreted as a mixture between the reference microphone signals (scaled with $\eta$), the binauralized output signals of the D-BLCMV (scaled with $1-\eta$) and the binauralized output signals of the I-BLCMV (scaled with $\delta - \eta$).

Due to the constraints in \eqref{eq:costBLCMVN}, the BLCMV-N perfectly preserves the desired source component and scales the interfering source component with $\delta$.
%
Using \eqref{eq:decomp2} and \eqref{eq:decomp2R}, the noise component in the output signals of the BLCMV-N are equal to
\begin{IEEEeqnarray}{rCl}
\label{eq:noiseOutput}
	\wLblcmvn^H\vecn &=& \eta n_L + (1-\eta)n_{x} a_L + (\delta - \eta) n_{u} b_L \,, \IEEEeqnarraynumspace \\
	\wRblcmvn^H\vecn &=& \eta n_R + (1-\eta)n_{x} a_R + (\delta - \eta) n_{u} b_R \,, \IEEEeqnarraynumspace
\end{IEEEeqnarray}
with $n_{x} = \wx^H\vecn$ and $n_{u} = \wu^H\vecn$ the noise component in the output signal of the D-BLCMV and the I-BLCMV, respectively.
The noise component in the output signals of the BLCMV-N can hence be interpreted as a mixture between the noise component in the reference microphone signals (scaled with $\eta$), a coherent residual noise source ($n_{x}$) coming from the direction of the desired source (scaled with $1 - \eta$) and a coherent residual noise source ($n_{u}$) coming from the direction of the interfering source (scaled with $\delta - \eta$).
\section{Performance of the BLCMV-N}
\label{sec:performance}
In this section we provide a performance analysis of the proposed BLCMV-N.
In Section~\ref{sec:lcmvnpsds} we derive the output PSDs of the signal components.
In Sections~\ref{sec:lcmvnnrperformance} and \ref{sec:lcmvnbinauralcues} we analyze the noise and interference reduction performance and the binaural cue preservation performance.
Finally, in Section~\ref{sec:lcmvnscalingparametersetting} we discuss the setting of the mixing parameter $\eta$ and the interference scaling parameter $\delta$.
\subsection{Output Power Spectral Densities}
\label{sec:lcmvnpsds}
Due to the constraints in \eqref{eq:costBLCMVN}, the output PSD of the desired and interfering source components in the left output signal of the BLCMV-N are equal to, cf. \eqref{eq: PSDin},
\begin{IEEEeqnarray}{rCl}
	\label{eq: psdoutx}
	p^{\mathrm{out}}_{\mathrm{BLCMV-N},x,L} &=& p^{\mathrm{in}}_{x,L} = p_x|a\links|^2  \, , \\
	\label{eq: psdoutu}
	p^{\mathrm{out}}_{\mathrm{BLCMV-N},u,L} &=& \delta^2 p^{\mathrm{in}}_{u,L} = \delta^2 p_u |b\links|^2 \, .
\end{IEEEeqnarray}
Furthermore, the output PSD of the noise component in the left output signal of the BLCMV-N is equal to (see Appendix~\ref{app2})
\begin{equation}
	\label{eq:psdnout}
	p^{\mathrm{out}}_{\mathrm{BLCMV-N},n,L} = \veceL^T \left( \eta^2 \matRn + \mathbf{R}_{xu,3} \right) \veceL \, ,
\end{equation}
with
\begin{IEEEeqnarray}{rCl}
	\label{eq:Rxu}
		\mathbf{R}_{xu,3} &=& \frac{1}{1 - \Psi} \left[( 1 -\eta^2) \frac{\veca\veca^H}{\ga} + (\delta^2 - \eta^2) \frac{\vecb\vecb^H}{\gb}\right.\\
		&&\left. - 2\Psi(\delta - \eta^2) \Re\left\{\frac{\veca\vecb^H}{\gab^*}\right\}\right]\nonumber \, ,
\end{IEEEeqnarray}
with $\gamma_a$ defined in \eqref{eq: gamma_a}, $\gamma_{ab}$ defined in \eqref{eq: gamma_ab}, and $\gamma_b$ and $\Psi$ defined in \eqref{eq: gamma_b_Psi}.
It can be seen that the output PSD of the noise component for the BLCMV-N is a quadratic function in both the mixing parameter $\eta$ and the interference scaling parameter $\delta$.
By comparing \eqref{eq:Rxu} to \eqref{eq: Rxu1}, it can be observed that
\begin{IEEEeqnarray}{C}
	\label{eq: relation}
	\boxed{
	\mathbf{R}_{xu,3} = \mathbf{R}_{xu,1} - \eta^2\mathbf{R}_{xu,1}^{\delta=1} }
\end{IEEEeqnarray}
where $\mathbf{R}_{xu,1}^{\delta=1}$ denotes the expression for the BLCMV in \eqref{eq: Rxu1} with $\delta=1$, corresponding to no suppression of the interfering source.
Please note that for $\eta=0$, $\mathbf{R}_{xu,3} = \mathbf{R}_{xu,1}$, and for $\eta=1$ and $\delta=1$, $\mathbf{R}_{xu,3} = \mathbf{0}_M$.
By using \eqref{eq: relation} in \eqref{eq:psdnout}, it follows that
\begin{equation}
	\label{eq: simonsEquality}
	p^{\mathrm{out}}_{\mathrm{BLCMV-N},n,L} = \eta^2 \left( p_{n,L}^\mathrm{in} - p^{\mathrm{out},\delta=1}_{\mathrm{BLCMV},n,L} \right) + p^{\mathrm{out}}_{\mathrm{BLCMV},n,L} \, .
\end{equation}

%
%
%
\subsection{Noise and Interference Reduction Performance}
\label{sec:lcmvnnrperformance}
By substituting \eqref{eq: psdoutx} and \eqref{eq:psdnout} in \eqref{eq:SNRout}, the left output SNR of the BLCMV-N is equal to
\begin{equation}
	\label{eq:lcmvnSNRout}
	\SNRoutLlcmvn = \frac{p_x |a\links|^2}{\veceL^T(\eta^2\matRn + \mathbf{R}_{xu,3})\veceL} \, ,
\end{equation}
which depends on both the mixing parameter $\eta$ and the interference scaling parameter $\delta$.
Using \eqref{eq: simonsEquality} and realizing that the output PSD of the noise component in the left output signal of the BLCMV (for any value of $\delta$) is smaller than or equal to the PSD of the noise component in the left reference microphone signal, the output SNR of the BLCMV-N in \eqref{eq:lcmvnSNRout} is smaller than or equal to the output SNR of the BLCMV in \eqref{eq: outSNR_blcmv_L}, i.e.,
\begin{IEEEeqnarray}{C}
	\label{eq: SNRproof}
	\boxed{
		\mathrm{SNR}_{\mathrm{BLCMV-N},L}^\mathrm{out} \leq \mathrm{SNR}_{\mathrm{BLCMV},L}^\mathrm{out} \leq \mathrm{SNR}_{\mathrm{BMVDR},L}^\mathrm{out}} \IEEEeqnarraynumspace
\end{IEEEeqnarray}
By substituting \eqref{eq: psdoutx} and \eqref{eq: psdoutu} in \eqref{eq:SIRout}, the left output SIR of the BLCMV-N is equal to
\begin{equation}
	\label{eq:lcmvnSIRout}
	\SIRoutLlcmvn = \frac{1}{\delta^2} \SIRinL \, ,
\end{equation}
which is equal to the left output SIR of the BLCMV in \eqref{eq:lcmvSIRout} and solely controlled by the interference scaling parameter $\delta$.
For $\eta = 0$, the left output SNR of the BLCMV-N is equal to the left output SNR of the BLCMV in \eqref{eq: outSNR_blcmv_L}, while for $\eta = 1$ and $\delta = 1$, the left output SNR of the BLCMV-N is equal to the left input SNR because no beamforming is applied.
%
%
\subsection{Binaural Cue Preservation}
\label{sec:lcmvnbinauralcues}
Similarly as for the BLCMV, due to the constraints in \eqref{eq:costBLCMVN} the BLCMV-N preserves the binaural cues of both the desired source and the interfering source, i.e.,
\begin{IEEEeqnarray}{rClCl}
	\mathrm{ITF}_{\mathrm{BLCMV-N},x}^{\mathrm{out}} &=& \frac{a\links}{a\rechts} &=& \mathrm{ITF}_{{x}}^{\mathrm{in}}\, ,\\
	\mathrm{ITF}_{\mathrm{BLCMV-N},u}^{\mathrm{out}} &=& \frac{b\links}{b\rechts} &=& \mathrm{ITF}_{{u}}^{\mathrm{in}}\, .
\end{IEEEeqnarray}
Using \eqref{eq: outICn}, the output IC of the noise component for the BLCMV-N is equal to (see Appendix~\ref{app2} for derivation of components)
\begin{IEEEeqnarray}{l}
	\label{eq:lcmvnicnout}
		\mathrm{IC}_{\mathrm{BLCMV-N},n}^\mathrm{out} = \nonumber \\
		\frac{\veceL^T(\eta^2\matRn + \mathbf{R}_{xu,3})\veceR}{\sqrt{\veceL^T(\eta^2\matRn + \mathbf{R}_{xu,3})\veceL} \sqrt{\veceR^T(\eta^2\matRn + \mathbf{R}_{xu,3})\veceR}} \, ,
\end{IEEEeqnarray}
with $\mathbf{R}_{xu,3}$ defined in \eqref{eq:Rxu}.
Since $\mathbf{R}_{xu,3}$ depends on both the mixing parameter $\eta$ and the interference scaling parameter $\delta$, also the output IC of the noise component in \eqref{eq:lcmvnicnout} depends on both parameters.
Using \eqref{eq:MSCout}, the output MSC of the noise component for the BLCMV-N is equal to
\begin{equation}
	\mathrm{MSC}_{\mathrm{BLCMV-N},n}^{\mathrm{out}} = |\mathrm{IC}_{\mathrm{BLCMV-N},n}^\mathrm{out}|^2 .
\end{equation}
Since for $\eta=0$ the BLCMV-N is equal to the BLCMV, the output MSC of the noise component is smaller than 1, see Section~\ref{sec:lcmv}.
%
%
It should however be realized that in contrast to the BMVDR-N discussed in Section~\ref{sec:mvdrn}, for $\eta=1$ the BLCMV-N does not always preserve the MSC of the noise component.
Only for $\eta=1$ and $\delta=1$ the binaural cues of all signal components are preserved because no beamforming is applied.
%
%
\subsection{Parameter Settings}
\label{sec:lcmvnscalingparametersetting}
Maximizing the left output SNR in \eqref{eq:lcmvnSNRout} corresponds to minimizing the denominator, i.e., using \eqref{eq: relation},
\begin{IEEEeqnarray}{C}
	\label{eq: denom}
	D(\eta,\delta) = \mathbf{e}_L^T \left[ \eta^2 \left(\mathbf{R}_n - \mathbf{R}_{xu,1}^{\delta=1} \right) + \mathbf{R}_{xu,1} \right] \mathbf{e}_L \, .
\end{IEEEeqnarray}
Setting the derivative of \eqref{eq: denom} with respect to the mixing parameter $\eta$ equal to zero, yields
\begin{equation}
	\eta_{\mathrm{opt}} = 0
\end{equation}
as the optimal mixing parameter $\eta$ in terms of left (and right) output SNR.
The derivative of \eqref{eq: denom} with respect to the interference scaling parameter $\delta$ is equal to, using \eqref{eq: Rxu1},
\begin{equation}
	\label{eq: deriv2}
	\frac{\partial D(\eta,\delta)}{\partial\delta} = \frac{1}{1- \Psi} \left( 2\delta \frac{|b\links|^2}{\gamma_b} - 2 \Psi \Re\left\{ \frac{a_L b_L^*}{\gamma_{ab}^*} \right\} \right) \, .
\end{equation}
Setting \eqref{eq: deriv2} to zero and solving for $\delta$ yields the optimal interference scaling parameter in terms of left output SNR, i.e.,
\begin{equation}
	\delta_{\mathrm{opt},L} = \frac{\alpha\links}{\beta\links} \, ,
	\label{eq:deltaOPT}
\end{equation}
with
\begin{equation}
	\alpha\links = \Psi \Re\left\{\frac{a\links b\links^*}{\gab^*}\right\} \, , \quad \beta\links = \frac{|b\links|^2}{\gb} \, .
\end{equation}
As can be seen from \eqref{eq:lcmvnSIRout}, the output SIR is not affected by the mixing parameter $\eta$ but is solely determined by the interference scaling parameter $\delta$.
%
%
%
\section{Simulations}
\label{sec:simulations}
In Section~\ref{subsec:validation} we first validate the expressions derived in the previous sections using measured anechoic ATFs.
In Section~\ref{sec:Exp} we then experimentally compare the performance of the proposed BLCMV-N with the BMVDR, BLCMV and BMVDR-N using recorded signals in a reverberant environment with a competing speaker and multi-talker babble noise.
Finally, in Section~\ref{subsec:listeningTest} we compare the spatial impression of the considered binaural beamforming algorithms using a perceptual listening test.
%
\subsection{Validation Using Measured Anechoic ATFs}
\label{subsec:validation}
To validate the derived expressions for the considered algorithms we used measured anechoic ATFs of two behind-the-ear hearing aids mounted on a head-and-torso-simulator (HATS) \cite{Kayser2009}.
Each hearing aid has two microphones ($M=4$) with an inter-microphone distance of about \SI{14}{\milli\metre}.
We chose the front microphone on each hearing aid as reference microphone.
The ATFs were calculated from anechoic impulse responses using a 512-point FFT at a sampling rate of \SI{16}{\kilo\hertz}.

The desired source was placed at \SI{0}{\degree} (in front) and the interfering source was placed at \SI{-35}{\degree} (to the left), both at a distance of \SI{3}{\metre} from the HATS.
The desired source covariance matrix $\matRx$ and the interfering source covariance matrix $\matRu$ were constructed using the ATF vector of the desired source $\mathbf{a}$ and the ATF vector of the interfering source $\mathbf{b}$ according to \eqref{eq:rank1assum}, where the PSD of the desired source $p_x$ and the PSD of the interfering source $p_u$ were both set to 1.
As background noise we considered a combination of spatially white and cylindrically isotropic noise, i.e., the noise covariance matrix $\matRn$ was constructed as
\begin{equation} \label{eq:Rn_simu}
	\matRn = p_{n,\mathrm{w}} \mathbf{I}_M + p_{n,\mathrm{cyl}} \bm{\Gamma} \, ,
\end{equation}
with $p_{n,\mathrm{w}}$ the PSD of the spatially white noise, $\mathbf{I}_M$ the $M \times M$-dimensional identity matrix, $p_{n,\mathrm{cyl}}$ the PSD of the cylindrically isotropic noise and $\bm{\Gamma}$ its spatial coherence matrix.
The $(i,j)$-th element of the spatial coherence matrix $\bm{\Gamma}$ was calculated using all available anechoic ATFs as
\begin{equation}
	\bm{\Gamma}_{i,j} = \frac{\sum^K_{k=1}h_i(\theta_k)h^*_j(\theta_k)}{\sqrt{ \sum^K_{k=1}|h_i(\theta_k)|^2} \sqrt{ \sum^K_{k=1}|h_j(\theta_k)|^2}} \, ,
\end{equation}
with $h(\theta_k)$ the anechoic ATF at angle $\theta_k$ and $K$ the total number of angles in the database ($K=72$ for \cite{Kayser2009}).
The PSD of the spatially white noise $p_{n,\mathrm{w}}$ was set to \SI{-55}{\decibel}, while the PSD of the cylindrically isotropic noise $p_{n,\mathrm{cyl}}$ was set to 1.
%
%
\subsubsection{Noise and Interference Reduction Performance}
\begin{figure}[t]
	\centering
	\includegraphics[scale=0.9]{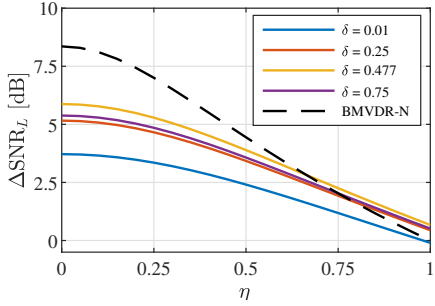}
	\vspace{-3mm}
	\caption{SNR improvement for the BLCMV-N and the BMVDR-N at \SI{500}{\hertz}.}
	\label{fig: dSNR_vali}
\end{figure}
\begin{figure}[t]
	\centering
	\includegraphics[scale=0.9]{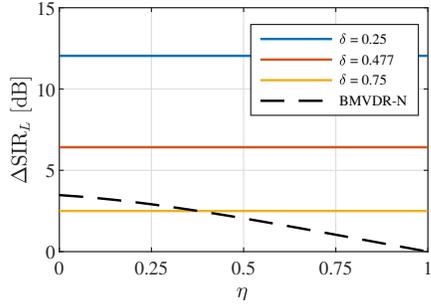}
	\vspace{-3mm}
	\caption{SIR improvement for the BLCMV-N and the BMVDR-N at \SI{500}{\hertz}.}
	\label{fig: dSIR_vali}
\end{figure}
Using \eqref{eq: SNRin} and \eqref{eq:SNRout}, Figure \ref{fig: dSNR_vali} depicts the left SNR improvement at \SI{500}{\hertz} for the BLCMV-N for different values of the mixing parameter $\eta$ and the interference scaling parameter $\delta$ and the BMVDR-N for different values of the mixing parameter $\eta$.
As expected, the BMVDR (i.e., BMVDR-N for $\eta=0$) yields the largest SNR improvement (cf. \eqref{eq: SNRproof}).
Since the BMVDR-N mixes the output signals of the BMVDR with the noisy reference microphone signals, it can be observed that increasing the mixing parameter $\eta$ reduces the SNR improvement of the BMVDR-N compared to the BMVDR ($\eta=0$).
For the BLCMV-N, both $\eta$ and $\delta$ affect the SNR improvement, which is in line with \eqref{eq:lcmvnSNRout}.
Similarly to the BMVDR-N, the BLCMV-N mixes the output signals of a BLCMV with the noisy reference microphone signals.
Hence, it can be observed that for any value of the interference scaling parameter $\delta$, increasing the mixing parameter $\eta$ reduces the SNR improvement of the BLCMV-N compared to the BLCMV ($\eta=0$), which is in line with \eqref{eq: SNRproof}.
Since less degrees of freedom are available for noise reduction, the BLCMV ($\eta=0$) yields a smaller SNR improvement compared to the BMVDR ($\eta=0$), as discussed in Section~\ref{sec:lcmv}.
Using \eqref{eq:deltaOPT}, the interference scaling parameter $\delta$ maximizing the output SNR was equal to $\delta_{\mathrm{opt},L} = 0.477$ for the considered acoustic scenario.
As expected, it can be observed that using $\delta_{\mathrm{opt},L}$ leads to the largest SNR improvement of all considered values of $\delta$.
For large values of the mixing parameter $\eta$, the BLCMV-N yields a larger SNR improvement than the BMVDR-N.
It should be noted that the exact behaviour depends on the interference scaling parameter $\delta$ and the relative position of the interfering source to the desired source.

Using \eqref{eq: SIRin} and \eqref{eq:SIRout}, Figure \ref{fig: dSIR_vali} depicts the left SIR improvement at \SI{500}{\hertz} for the BLCMV-N for different values of the mixing parameter $\eta$ and the interference scaling parameter $\delta$ and the BMVDR-N for different values of the mixing parameter $\eta$.
As expected from \eqref{eq:lcmvSIRout} and \eqref{eq:lcmvnSIRout}, both the BLCMV-N and the BLCMV ($\eta=0$) yield the same SIR improvement, which is solely controlled by the interference scaling parameter $\delta$.
Hence, increasing the interference scaling parameter $\delta$ reduces the SIR improvement for both the BLCMV-N and the BLCMV.
For the BMVDR-N it can be observed that increasing the mixing parameter $\eta$ reduces the SIR improvement.
It should be noted that the exact behaviour depends on the relative position of the interfering source to the desired source, as can be seen from \eqref{eq: outSIR_bmvdrn_L} and \eqref{eq: Rxu2}.
%
%
\subsubsection{Binaural Cue Preservation of Background Noise}
\begin{figure}[t]
	\centering
	\includegraphics[scale=0.9]{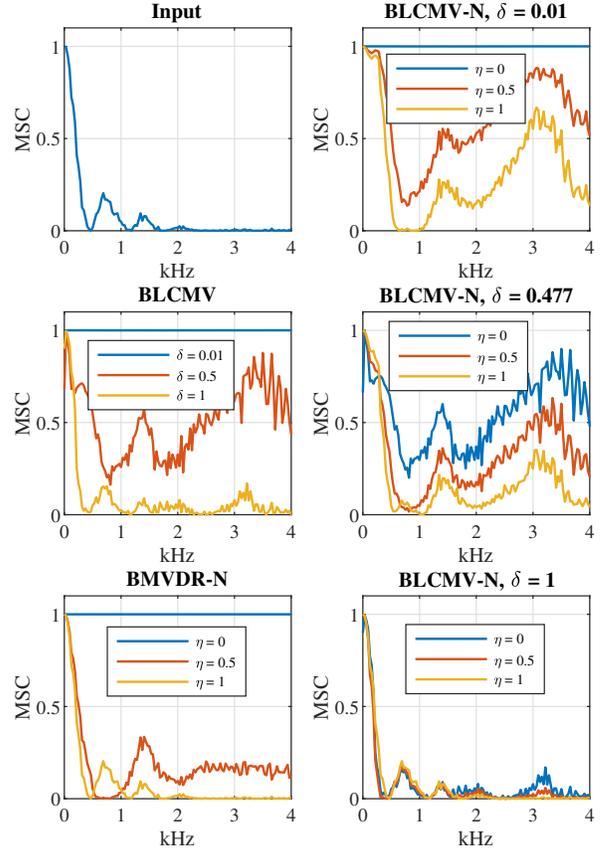}
	\vspace{-2mm}
	\caption{The MSC of the noise component in the reference microphone signals (\textbf{Input}), in the output signals of the BLCMV for different values of the interference scaling parameter $\delta$, the BMVDR-N for different values of the mixing parameter $\eta$ and the BLCMV-N for different values of the mixing parameter $\eta$ and the interference scaling paramter $\delta$.}
	\label{fig:MSCno}
\end{figure}
For different frequencies, Figure \ref{fig:MSCno} depicts the input MSC in \eqref{eq:MSCout} of the noise component (\textbf{Input}) and the output MSC in \eqref{eq:MSCout} of the noise component for the BLCMV in \eqref{eq: ICn_blcmv} for different values of the interference scaling parameter $\delta$, the BMVDR-N in \eqref{eq: outMSC_bmvdrn} for different values of the mixing parameter $\eta$ and the BLCMV-N for different values of the mixing parameter $\eta$ and the interference scaling parameter $\delta$.
Although the BLCMV is not designed to preserve the MSC of the noise component, it can be observed that an output MSC smaller than 1 is obtained, especially for large values of $\delta$ \cite{Hadad2016}.
However, since the output MSC of the noise component depends on the relative position of the interfering source to the desired source, it cannot be easily controlled.
Since the BMVDR-N mixes the output signals of the BMVDR with the noisy reference microphone signals, it can be observed that the output MSC of the noise component is smaller than 1, and for $\eta=1$ the MSC is perfectly preserved (but no beamforming is applied).
For the BLCMV-N, it can be observed that both $\eta$ and $\delta$ influence the output MSC of the noise component, as discussed in Section~\ref{sec:lcmvnbinauralcues}. 
For $\eta=0$, the output MSC of the noise component for the BLCMV-N is obviously equal to the output MSC of the noise component for the BLCMV.
For a fixed value of $\delta$, it can be observed that the output MSC of the noise component approaches the input MSC of the noise component for increasing $\eta$, although it should be realized that perfect preservation of the MSC of the noise component is only possible for $\delta = 1$ (cf. Section~\ref{sec:lcmvnbinauralcues}).

\begin{figure}[t]
	\centering
	\includegraphics[scale=0.9]{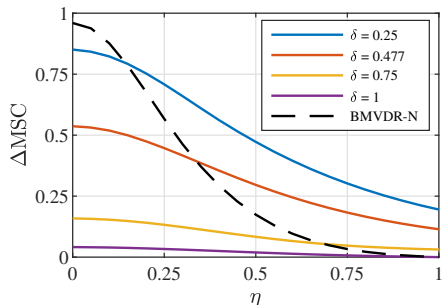}
	\vspace{-3mm}
	\caption{Frequency-averaged MSC error of the noise component for the BLCMV-N and the BMVDR-N.}
	\label{fig: dMSCn_intersec}
\end{figure}
For several values of the mixing parameter $\eta$, Figure \ref{fig: dMSCn_intersec} depicts the MSC error of the noise component for the BLCMV-N and the BMVDR-N, averaged over all frequencies, i.e.,
\begin{equation}
	\label{eq: MSCerror}
	\Delta\mathrm{MSC} = \frac{1}{F-1} \sum_{f=1}^{F-1}|\mathrm{MSC}^{\mathrm{in}}_{n}(f) - \mathrm{MSC}^{\mathrm{out}}_n(f)|\, ,
\end{equation}
with $f$ the frequency bin index and $F$ the total number of frequency bins.
As expected, the BMVDR ($\eta=0$) yields the largest MSC error of the noise component and increasing the mixing parameter $\eta$ reduces the frequency-averaged MSC error of the noise component for the BMVDR-N \cite{Marquardt2018}.
For the considered acoustic scenario, it can be observed for the BLCMV-N that for any value of the interference scaling parameter $\delta$, increasing the mixing parameter $\eta$ reduces the frequency-averaged MSC error of the noise component compared to the BLCMV ($\eta=0$).
Further, it can be observed that for small values of the interference scaling parameter $\delta$, the effect of the mixing parameter $\eta$ is larger than for large values of the interference scaling parameter $\delta$, for which the frequency-averaged MSC error is relatively small for all values of the mixing parameter $\eta$.
These results clearly show that the mixing parameter $\eta$ in the BLCMV-N enables to control the binaural cues of the background noise.
%
%
%
%
\subsection{Experimental Results Using Reverberant Recordings}
\label{sec:Exp}
For a more realistic evaluation, we compare the performance of the considered binaural beamforming algorithms using reverberant recordings.
Similarly to Section~\ref{subsec:validation}, the experimental setup consists of two hearing aids, each with two microphones, mounted on a HATS in a cafeteria with a reverberation time of approximately \SI{1.25}{\second} \cite{Kayser2009}.
The desired source was again placed at \SI{0}{\degree} (at a distance of about \SI{102}{\centi\metre}), while the interfering source was again placed at \SI{-35}{\degree} (at a distance of about \SI{118}{\centi\metre}), see \cite{Kayser2009} for more details.
The desired and interfering source components were generated by convolving clean speech signals with the measured reverberant room impulse responses corresponding to the desired source and interfering source positions.
The desired source was a male German speaker, speaking eight sentences with a pause of \SI{1}{\second} between the sentences.
The interfering source was a male Dutch speaker, speaking seven sentences with a pause of \SI{0.25}{\second} between the sentences.
As background noise we used realistic recordings \cite{Kayser2009}, consisting of multi-talker babble noise, clacking plates and temporally dominant competing speakers.
The used background noise hence clearly differed from the perfectly diffuse noise in Section~\ref{subsec:validation}.
The entire signal had a length of about \SI{28}{\second}.
The desired source and the background noise were active the entire time, whereas the interfering source only became active after about \SI{14}{\second}.
The desired source component, the interfering source component and the noise component were mixed at an input SNR of \SI{10}{\decibel} and input SIR of \SI{5}{\decibel} in the right reference microphone.
Again, we chose the front microphone on each hearing aid as reference microphone.

The processing was performed at a sampling rate of \SI{16}{\kilo\hertz} in the STFT domain with a frame length of \num{8192} samples and a square-root Hann window with \SI{50}{\percent} overlap.
We used an oracle voice activity detector (i.e., using the desired source and interfering source signals) to estimate the noise covariance matrix $\matRn$, the undesired covariance matrix $\matRv$ (interfering source plus background noise) and $\mathbf{R}_{xn} = \matRx + \matRn$ (desired source plus background noise) over the entire signal.
All binaural beamforming algorithms were implemented using relative transfer function (RTF) vectors \cite{Gannot2001}, relating the ATF vectors in \eqref{eq:atfSchreibweise} to the reference microphones.
Using the covariance whitening method (see \cite{Hadad2016,Markovich2009} for further details) the RTF vectors of the desired source and the interfering source were estimated based on generalised eigenvalue decomposition of $\mathbf{R}_{xn}$ and $\matRn$ or $\matRv$ and $\matRn$, respectively.
The mixing parameter was set to $\eta = 0.3$ and the interference scaling parameter was set to $\delta=0.3$.

As objective performance measures for noise and interference reduction performance, we used the left and the right SNR improvement ($\Delta\mathrm{SNR}_{L}$, $\Delta\mathrm{SNR}_{R}$) and the left and the right SIR improvement ($\Delta\mathrm{SIR}_{L}$, $\Delta\mathrm{SIR}_{R}$).
As objective performance measure for binaural cue preservation of the background noise we used the frequency-averaged MSC error of the noise component ($\Delta\mathrm{MSC}$) as defined in \eqref{eq: MSCerror}.
All objective performance measures were computed using the reference microphone signals and the output signals of all considered algorithms. 
Table \ref{tab:table1} presents the objective performance measures for all considered algorithms.
\begin{table}
	\centering
	\caption{Objective performance measures for all considered algorithms in the reverberant environment.}
	\begin{tabular}{|l|c|c|c|c|}
		\hline
		 & BMVDR & BLCMV & BMVDR-N & BLCMV-N\\
		 \hline
		 $\Delta\mathrm{SNR}_{L}$ [dB] & 13.0 & 10.1 & 8.6 & 7.6 \\ \hline
		 $\Delta\mathrm{SNR}_{R}$ [dB] & 12.9 & 9.2 & 8.6 & 7.0 \\ \hline
		 $\Delta\mathrm{SIR}_{L}$ [dB] & -0.1 & 9.7 & 0.82 & 9.8 \\ \hline
		 $\Delta\mathrm{SIR}_{R}$ [dB] & -4.3 & 8.7 & -2.4 & 8.9 \\ \hline
		 $\Delta\mathrm{MSC}$ & 0.86 & 0.64 & 0.10 & 0.19 \\ \hline
	\end{tabular}
	\label{tab:table1}
\end{table}
\begin{figure*}[ht]
	\centering
	\includegraphics[scale=0.9]{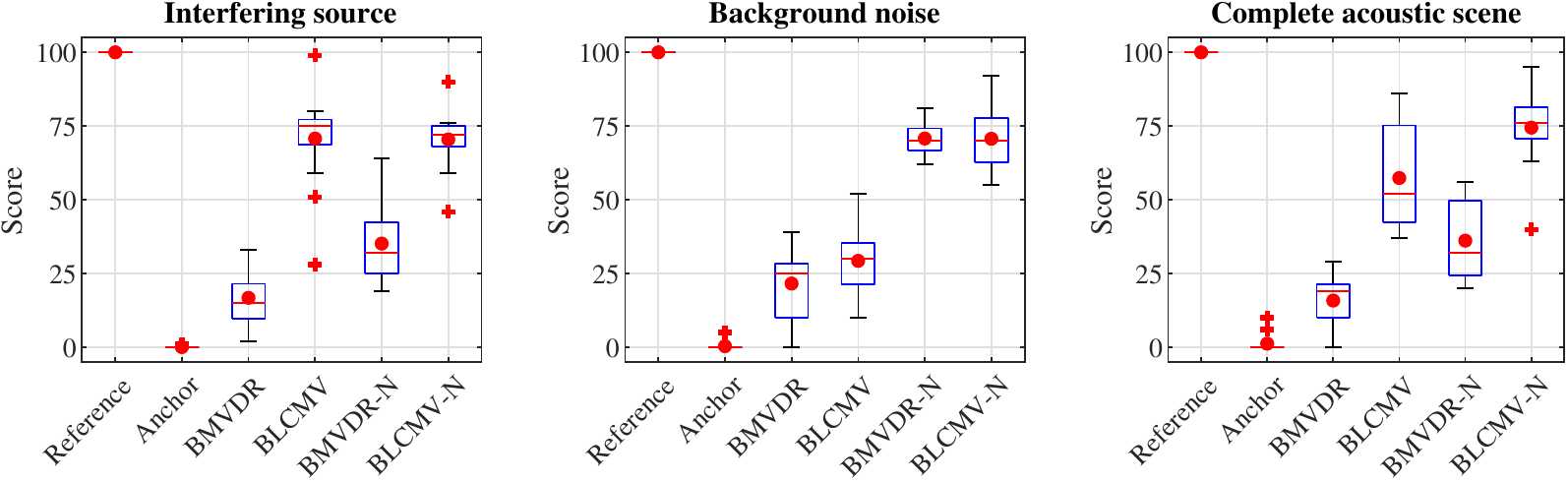}
	\vspace{-3mm}
	\caption{Boxplot of the MUSHRA scores for all three evaluations. The plot depicts the median score (red line), the mean score (red dot), the first and third quartiles (blue boxes) and the interquartile ranges (whiskers). Outliers are indicated by red $+$ markers.}
	\label{fig:mushra}
\end{figure*}
%
%
\subsubsection{Noise and Interference Reduction Performance}
In terms of noise reduction performance, it can be observed that -- as expected -- the BMVDR yields the highest SNR improvement (\SI{13.0}{\decibel} for the left and \SI{12.9}{\decibel} for the right side).
All other algorithms yield a lower SNR improvement, for the BLCMV due to the additional constraint for the interfering source, for the BMVDR-N due to the mixing with the noisy reference microphone signals, and for the BLCMV-N due to both effects.
The partial noise estimation for the BLCMV-N seems to result in a smaller drop in noise reduction performance compared to the BLCMV (\SI{2.5}{\decibel} for the left side, \SI{2.2}{\decibel} for the right side) than for the BMVDR-N compared to the BMVDR (\SI{4.4}{\decibel} for the left side, \SI{4.3}{\decibel} for the right side).
Please note that both for the BMVDR-N as well as for the BLCMV-N this drop in noise reduction performance depends on the relative position of the interfering source to the desired source.

In terms of interference reduction performance, it can be observed that both the BLCMV and the BLCMV-N approximately lead to the same SIR improvement (for the left and the right side), which is in line with the theoretical SIR improvement in \eqref{eq:lcmvSIRout} and \eqref{eq:lcmvnSIRout}, i.e., $10\log_{10} \frac{1}{\delta^2} \approx \SI{10.5}{\decibel}$.
The fact that this theoretical SIR improvement is not reached and the fact that the SIR improvements for the BLCMV and BLCMV-N are not exactly the same is due to estimation errors in the covariance matrices, which was also already noted in \cite{Hadad2016,Goessling2018_iwaenc_b}.
In addition, it can be observed that the BMVDR and BMVDR-N lead to very low (even negative) SIR improvements, which is presumably due to the fact that the interfering source is relatively close to the desired source.
%
%
\subsubsection{Binaural Cue Preservation of Background Noise}
As expected, the BMVDR yields the largest MSC error of the noise component $\Delta\mathrm{MSC}$.
As discussed in Section \ref{sec:lcmv}, the output MSC of the noise component for the BLCMV is typically smaller than 1, hence leading to a smaller MSC error compared to the BMVDR.
Due to the mixing with the noisy reference microphone signals, both the BMVDR-N and the BLCMV-N yield a much smaller MSC error of the noise component than the BMVDR and the BLCMV, where the MSC error is slightly smaller for the BMVDR-N than for the BLCMV-N.

In conclusion, the objective performance measures show that the BLCMV-N leads to a very similar interference reduction as the BLCMV, while providing a trade-off between noise reduction performance (slightly worse than the BLCMV) and binaural cue preservation of the background noise (much better than the BLCMV).
%
%
%
%
\subsection{Perceptual Listenting Test}
\label{subsec:listeningTest}
To further investigate the spatial impression of the different output signal components for the four considered algorithms, we conducted a perceptual listening test similarly to \cite{Goessling2020_tih}.
The desired source was now placed at \SI{-35}{\degree} and the interfering source was placed at \SI{90}{\degree}, in order to enhance the perceived spatial differences between both sources.
The desired source component, the interfering source component and the noise component were mixed at an input SNR of \SI{0}{\decibel} and input SIR of \SI{0}{\decibel} in the right reference microphone.
Thirteen self-reported normal-hearing subjects participated in the perceptual listening test, where none of the authors participated.
All subjects can be considered expert listeners, i.e., they were familiar with similar perceptual listening tests, and gave informed consent.
The listening test was conducted in a sound proof listening booth using an RME Fireface UCX sound card with Sennheiser HD 580 headphones.

Using a procedure similar to the MUlti-Stimulus Test with Hidden Reference and Anchor (MUSHRA) \cite{Mushra}, the task was to rate the perceived spatial difference with respect to a reference signal.
For a coherent source (e.g., interfering source), this corresponds to rating differences in perceived source location, whereas for a diffuse noise field this corresponds to rating differences in perceived diffuseness.
A score of 0 is associated with a large perceived spatial difference, whereas a score of 100 is associated with no perceived spatial difference.
As reference signal we used the (unprocessed) reference microphone signals, while as anchor signal we used the left reference microphone signal, played back to both ears.
The anchor signal was hence a monaural signal with no binaural cues, which is perceived in the center of the head.

We conducted three evaluations, where only some components were active in the output signals, the reference signal and the anchor signal.
In the first evaluation, only the desired source component and the interfering source component (i.e., no noise component) were active and the task was to rate the spatial difference for the interfering source.
In the second evaluation, only the desired source component and the noise component (i.e., no interfering source component) were active and the task was to rate the spatial difference for the background noise.
In the third evaluation, all signal components were active and the task was to rate the spatial difference for the interfering source and the background noise simultaneously.
To familiarize the subjects with the tasks and the sound material, a training round was performed.
Audio samples for all binaural beamforming algorithms and the unprocessed input signals are available online (see https://uol.de/en/sigproc/research/audio-demos/binaural-noise-reduction/blcmv-n-beamformer).

The MUSHRA scores for the three evaluations are shown in Figure \ref{fig:mushra}.
A one-way repeated-measures ANOVA was performed.
The analysis revealed a significant within-subjects effect for all three evaluations.
Hence, post-hoc comparison t-tests with Bonferroni correction were performed \cite{Kirkwood2010}.
\paragraph{Interfering source} The within-subjects effect was significant [$F(2.098,25.176)=219.2$, $p<.001$, Greenhouse-Geisser correction].
As expected, the BLCMV and the BLCMV-N preserved the spatial impression of the interfering source significantly better than the BMVDR and the BMVDR-N ($p < .001$).
The BMVDR-N performed significantly better than the BMVDR ($p < .001$), which is not unexpected since the interfering source component is also mixed with the mixing paremter $\eta$.
No significant difference was found between the BLCMV and the BLCMV-N ($p=1$).
\paragraph{Background noise} The within-subjects effect was significant [$F(3.072,36.869)=332.066$, $p<.001$, Greenhouse-Geisser correction].
As expected, the BMVDR-N and the BLCMV-N, both using partial noise estimation, preserved the spatial impression of the background noise significantly better than the BMVDR and the BLCMV ($p < .001$).
No significant difference was found between the BMVDR-N and the BLCMV-N ($p=1$) and between the BMVDR and BLCMV ($p=.614$).
\paragraph{Complete acoustic scene} The within-subjects effect was significant [$F(2.905,34.858)=171.783$, $p<.001$, Greenhouse-Geisser correction].
In terms of preservation of the spatial impression of the complete acoustic scene, the BMVDR-N scored significantly higher than the BMVDR ($p<.001$), the BLCMV scored significantly higher than the BMVDR-N ($p=.014$), and the proposed BLCMV-N scored significantly higher than the BLCMV ($p=.025$).

In summary, the results of the listening test showed that the BLCMV-N is capable of preserving the spatial impression of an interfering source and background noise in a realistic acoustic scenario, outperforming all other considered binaural beamforming algorithms in terms of spatial impression.
%
%
%
%
\section{Conclusions}
In this paper we proposed the BLCMV-N, merging the advantages of the BLCMV and the BMVDR-N, i.e., preserving the binaural cues of the interfering source and controlling the reduction of the interfering source as well as the binaural cues of the background noise.
We showed that the output signals of the BLCMV-N can be interpreted as a mixture between the noisy reference microphone signals and the output signals of a BLCMV using an adjusted interference scaling parameter.
We provided a theoretical comparison between the BMVDR, the BLCMV, the BMVDR-N and the proposed BLCMV-N in terms of noise and interference reduction performance and binaural cue preservation.
The obtained analytical expressions were first validated using measured anechoic acoustic transfer functions.
Experimental results using recorded signals in a realistic reverberant environment showed that the BLCMV-N leads to a very similar interference reduction as the BLCMV, while providing a trade-off between noise reduction performance (slightly worse than the BLCMV) and binaural cue preservation of the background noise (much better than the BLCMV).
In addition, the results of a perceptual listening test with 13 normal-hearing participants showed that the proposed BLCMV-N is capable of preserving the spatial impression of an interfering source and background noise in a realistic acoustic scenario, outperforming all other considered binaural beamforming algorithms in terms of spatial impression.
%
%
%
\appendices
%
\section{Derivation of the BLCMV-N}
\label{App1}
Using \eqref{eq:atfSchreibweise}, \eqref{eq: y_L} and \eqref{eq: CandgL}, the constrained optimization problem in \eqref{eq:costBLCMVN} can be reformulated as
\begin{equation}
	\min_{\wL}\mathcal{E}\left\{ \left| \wL^H\vecn - \eta n\links \right|^2\right\} \quad \text{s.t.} \quad \mathbf{C}^H\mathbf{w}_L = \mathbf{g}_L \, .
\end{equation}
This constrained optimization problem can be solved using the method of Lagrange multipliers, where the Lagrangian function is given by
\begin{IEEEeqnarray}{r}
	\mathcal{L}(\wL,\bm{\lambda}\links) = \wL^H\matRn\wL - \eta \veceL^T\matRn\wL - \eta \wL^H\matRn\veceL \IEEEeqnarraynumspace
 \\ 
	 + \eta^2 p_{n,L}^{\mathrm{in}} + \bm{\lambda}\links^H\left( \matC^H\wL - \vecgl \right) + \left( \wL^H\matC - \vecgl^H \right)\bm{\lambda}\links\, , \nonumber
\end{IEEEeqnarray}
with $\bm{\lambda}\links$ denoting the 2-dimensional vector of Lagrangian multipliers.
Setting the gradient with respect to $\wL$
\begin{equation} \label{eq:gradient}
	\nabla_{\wL}\mathcal{L}(\wL,\bm{\lambda}\links) = 2\matRn\wL - 2\eta\matRn\veceL + 2\matC\bm{\lambda}\links
\end{equation}
equal to $\mathbf{0}$ yields
\begin{equation} \label{eq:wBLCMVNderiv}
	\wL = \eta \veceL - \matRn^{-1}\matC\bm{\lambda}\links \, .
\end{equation}
Substituting \eqref{eq:wBLCMVNderiv} into the constraint $\mathbf{C}^H\mathbf{w}_L = \mathbf{g}_L$ and solving for the Lagrangian multiplier $\bm{\lambda}\links$ yields
\begin{equation} \label{eq:lagrangian}
	\bm{\lambda}\links = \left( \matC^H\matRn^{-1}\matC \right)^{-1} \left( \eta\matC^H\veceL - \vecgl \right) \, .
\end{equation}
Substituting \eqref{eq:lagrangian} into \eqref{eq:wBLCMVNderiv}, the solution to \eqref{eq:costBLCMVN} is given by
\begin{align}
	\wLblcmvn &= \\
	\eta \veceL &+ \matRn^{-1} \matC \left(\matC^H\matRn^{-1}\matC\right)^{-1} \left(\vecgl - \eta \matC^H\veceL\right) \, , \nonumber
\end{align}
where, using \eqref{eq: CandgL},
\begin{equation}
	\mathbf{g}_L - \eta \mathbf{C}^H \mathbf{e}_L = \begin{bmatrix}
		(1-\eta) a_L^* \\
		(\delta - \eta) b_L^*
	\end{bmatrix} \, .
\end{equation}
%
%
%
\section{Output noise PSD for the BLCMV-N}
\label{app2}
Using \eqref{eq:decomp2} in \eqref{eq:psdout} with $\mathbf{R}_n$ instead of $\mathbf{R}_x$, the output PSD of the noise component for the BLCMV-N is given by
\begin{IEEEeqnarray}{l}
	\label{eq: app2_1}
	\mathbf{w}_L^H \mathbf{R}_n \mathbf{w}_L = \eta^2 \mathbf{e}_L^T \mathbf{R}_n \mathbf{e}_L \\ \nonumber
	\quad + \eta(1-\eta) \left[ a_L \mathbf{w}_x^H \mathbf{R}_n \mathbf{e}_L + \mathbf{e}_L^T \mathbf{R}_n 	\mathbf{w}_x a_L^* \right] \\ \nonumber
	\quad + \eta(\delta-\eta) \left[b_L \mathbf{w}_u^H \mathbf{R}_n \mathbf{e}_L + \mathbf{e}_L^T \mathbf{R}_n 	\mathbf{w}_u b_L^* \right] \\ \nonumber
	\quad + (1-\eta)^2 \vert a_L \vert^2 \mathbf{w}_x^H \mathbf{R}_n \mathbf{w}_x \\ \nonumber
	\quad + (\delta-\eta)(1-\eta) \left[a_L^*b_L \mathbf{w}_u^H \mathbf{R}_n \mathbf{w}_x \right. + \left. a_L b_L^* \mathbf{w}_x^H \mathbf{R}_n \mathbf{w}_u \right] \\ \nonumber
	\quad + (\delta-\eta)^2 \vert b_L \vert^2 \mathbf{w}_u^H \mathbf{R}_n \mathbf{w}_u \, .
\end{IEEEeqnarray}
Using \eqref{eq:wx} and \eqref{eq:wu}, the components in \eqref{eq: app2_1} are given by \cite{Hadad2016}
\begin{IEEEeqnarray}{l}
	\label{eq: app2_2}
	\mathbf{e}_L^T \mathbf{R}_n \mathbf{w}_x = \frac{1}{1-\Psi} \left( \frac{a_L}{\gamma_a} - \Psi \frac{b_L}{\gamma_{ab}} \right), \,  \mathbf{w}_x^H \mathbf{R}_n \mathbf{w}_x = \frac{1}{(1-\Psi) \gamma_a}, \nonumber \\
	\mathbf{e}_L^T \mathbf{R}_n \mathbf{w}_u = \frac{1}{1-\Psi} \left( \frac{b_L}{\gamma_b} - \Psi \frac{a_L}{\gamma_{ab}^*} \right), \,  \mathbf{w}_u^H \mathbf{R}_n \mathbf{w}_u = \frac{1}{(1-\Psi) \gamma_b}, \nonumber \\
	\mathbf{w}_x^H \mathbf{R}_n \mathbf{w}_u = \frac{\Psi}{(1-\Psi)\gamma_{ab}^*} \, .
\end{IEEEeqnarray}
Substituting \eqref{eq: app2_2} in \eqref{eq: app2_1} yields
\begin{IEEEeqnarray}{l}
	\label{eq: app2_3}
	\mathbf{w}_L^H \mathbf{R}_n \mathbf{w}_L =\\
	\eta^2 p_{n,L}^{\mathrm{in}} + \frac{1}{1-\Psi} \left[ (1-\eta^2) \frac{\vert a_L \vert^2}{\gamma_a} + (\delta^2 - \eta^2) \frac{\vert b_L \vert^2}{\gamma_b} \right. \nonumber \\
	- \left. 2\Psi(\delta-\eta^2) \Re \left\{ \frac{a_L b_L^*}{\gamma_{ab}^*} \right\} \right] = \mathbf{e}_L^T ( \eta^2 \mathbf{R}_n + \mathbf{R}_{xu,3} ) \mathbf{e}_L \, , \nonumber
\end{IEEEeqnarray}
with $\mathbf{R}_{xu,3}$ defined in \eqref{eq:Rxu}.
Similarly, it can be shown that
\begin{IEEEeqnarray}{rCl}
	\label{eq: app3_1}
	\mathbf{w}_L^H \mathbf{R}_n \mathbf{w}_R &=& \mathbf{e}_L^T ( \eta^2 \mathbf{R}_n + \mathbf{R}_{xu,3} ) \mathbf{e}_R \, , \\
	\label{eq: app3_2}
	\mathbf{w}_R^H \mathbf{R}_n \mathbf{w}_R &=& \mathbf{e}_R^T ( \eta^2 \mathbf{R}_n + \mathbf{R}_{xu,3} ) \mathbf{e}_R \, .
\end{IEEEeqnarray}
%
%

\ifCLASSOPTIONcaptionsoff
  \newpage
\fi



\bibliographystyle{IEEEtran}
\bibliography{nicolib.bib}
%

%




\vfill


\end{document}